# Energy Aware Camera Location Search Algorithm for Increasing Precision of Observation in Automated Manufacturing

**Rongfei Li [1] and Francis Assadian [2]**

[1]  University of California, Davis; rfli@ucdavis.edu
[2]  University of California, Davis; fassadian@ucdavis.edu

**Abstract:** Visual servoing technology has been well developed and applied in many automated manufacturing tasks, especially in tools' pose alignment. To access a full global view of tools, most applications adopt eye-to-hand configuration or eye-to-hand/eye-in-hand cooperation configuration in an automated manufacturing environment. Most research papers mainly put efforts into developing control and observation architectures in various scenarios, but few of them have discussed the importance of the camera's location in eye-to-hand configuration. In a manufacturing environment, the quality of camera estimations may vary significantly from one observation location to another, as the combined effects of environmental conditions result in different noise levels of a single image shot at different locations. In this paper, we propose an algorithm for the camera's moving policy so that it explores the camera workspace and searches for the optimal location where the images' noise level is minimized. Also, this algorithm ensures the camera ends up at a suboptimal (if the optimal one is unreachable) location among the locations already searched, with limited energy available for moving the camera. Unlike a simple brute force approach, the algorithm enables the camera to explore space more efficiently by adapting the search policy from learning the environment. With the aid of an image averaging technique, this algorithm, in use of a solo camera, achieves the observation accuracy in eye-to-hand configurations to a desirable extent without filtering out high-frequency information in the original image. An automated manufacturing application has been simulated and the results show the success of this algorithm's improvement of observation precision with limited energy.

## 1. Introduction

In automated industry, automatic alignment plays a critical role in many different applications, such as micromanipulation, autonomous welding, and industrial assembly [1-3]. Visual servoing is the technique commonly used in the tasks of alignment as it can guide robotic manipulators to their desired poses (positions and orientations) [4-6]. Visual servoing is the method of controlling a robot's motion using real-time feedback from visual sensors that continuously extract image features [7].

According to the relative positions between cameras and the tool manipulators, visual servoing can be generally divided into two categories: eye-in-hand (EIH) configuration and eye-to-hand (ETH) configuration [8-9]. In EIH, a camera is mounted directly on a robot manipulator, in which case the motion of the robot induces motion of the camera, while in ETH, the camera is fixed in the workspace and observes the motion of the robot from a stationary configuration. Both configurations have their own merits and drawbacks regarding field of view limit and precision that oppose each other. EIH has a partial but precise view of the scene whereas ETH has a global but less precise view of it [10]. For complex tasks in automated manufacturing, one configuration of visual servoing is not adequate as those tasks require the camera to provide global views and to be maneuverable enough to explore the scene [10]. To take advantage of both stationary and robot-mounted sensors, two configurations can be employed in a cooperative scheme, which can be seen in some of the work [11-13].

In recent years, many works have been published to show visual servoing paradigm being applied in industrial environments. For example, Lippiello et al. [14] presented a position-based visual servoing approach for a multi-arm robotic cell equipped with a hybrid eye-in-hand/eye-to-hand multicamera system. Zhu et al. [15] discussed Abbe errors in a 2D vision system for robotic drilling. Four laser displacement sensors were used to improve the accuracy of the vison-based measurement system. Liu et al. [16,17] proposed a visual servoing method for automative assembly of aircraft components. With the measurements from two CCD cameras and four distances sensors, the proposed method can accurately align the ball-head which is fixed on the aircraft structures in a finite time. Sensor accuracy is a key element in many works as it is not only a requirement for some delicate applications but also affects the robustness of visual servoing techniques [18].

However, gaps exist between real-world applications and open-space experiments proposed in the above works. Real-world products are usually manufactured in a confined and complicated environment where complex factors may significantly degrade the quality of observed signals given from even the most precise sensors. For example, Illumination changes in space [18] could be a serious issue for camera's estimation in real manufacturing process, which is not normally addressed in research experiments where illumination is usually sufficient and equally distributed.

In this paper, we would like to address the importance of cameras' location in the ETH or ETH/EIH cooperative configurations and discuss the benefits of developing algorithms placing the camera in an optimal location. To observe a single point in the workspace, there are an infinite number of poses that a fixed camera can be placed in the space. Among all possible poses of the camera that keep the same target within the field of view limit, the image noise at different locations may play a great role in the precision of observations. Many environmental conditions (such as illumination, temperature, etc.) affect the noise level of a single image [19]. In a manufacturing environment, the quality of estimations from a camera may vary significantly from one observation location to another. For instance, the precision of an estimation improves greatly if the camera moves from a location where it is placed within the shade of machines to a location that has better illumination. Also, places near machines may be surrounded by strong electrical signals that also introduce extra noise into the camera sensors.

Many papers in visual servoing consider applying filters to reduce image noise, and therefore, to increase the precision of observations [20-23]. The Gaussian white noise can be dealt with using spatial filters, e.g., the Gaussian filter, the Mean filter, and the Wiener filter [24]. In addition, wavelet methods [25,26] have the benefit of keeping more useful details but at the expense of computational complexity. However, filters that operate in the wavelet domain still inevitably filter out (or blur) some important high-frequency useful information of the original image, even though they succeed in preserving more edges, compared with spatial filters. Especially at locations where the signal-to-noise ratio (SNR) in an image is low, no matter what filters are applied, it is difficult to safeguard the edges of the noise-free image as well as reduce the noise to a desirable level. Thus, developing an algorithm that searches for locations of the camera where SNR is high is beneficial.

We can approach the denoising problem with multiple noisy images taken from the same perspective. This method is called signal averaging (or image averaging in the application of image processing) [27]. Assume images taken from the same perspective but at different times have the same noise levels. Then averaging multiple images at the same perspective will reduce the unwanted noise as well as retain all original image details. We can also assume the robot's end effector holds the camera rigidly so that any shaking and drift are negligible when shooting pictures. Moreover, the number of images required for averaging has a quadratic dependence on the ratio of an image's original noise level to the reduced level that is desirable for observations. Therefore, the number of images required is a measurement of noise level in a single image. Furthermore, in the denoising process, the precise estimations require that the original image details are retained as much as possible. Based on previous statements, we decided to choose image averaging over all other denoising techniques in this work.

Frame-based cameras, which detect, track and match visual features by processing images at consecutive frames, cause a fundamental problem of delays in image processing and the consequent robot action [28]. By averaging more images for an observation, less noises are maintained in the averaged image, but more processing time is required for acquiring visual features; it is a typical tradeoff between high-rate and high-resolution in cameras [18]. The control loop frequency is limited in many cases due to low rate of imaging, which undermines the capability/usage of visual servoing technique in high-speed operations. Therefore, it is rewarding for an eye-in-hand configuration to find the location of the camera where original image noise is lowest and thus the least number of images is needed in order to facilitate the manufacturing speed.

Brute force search [29] is a simple and straightforward approach to look for the optimal location of the camera. It systematically searches all possible locations in space and gives the best solution. This method is only feasible when the size of search space is small enough. Otherwise, brute force which checks every possibility, is usually inefficient, especially for large datasets or complex problems, where the number of possibilities can grow exponentially. An intelligent algorithm is necessarily developed for the sake of energy efficiency and time consumption.

In this paper, we propose an algorithm that searches efficiently for the camera's workspace to find an optimal location (if its orientation is fixed) of the camera so that a single image taken at this location has the smallest noise level among images taken at all locations in the space. With limited energy for moving the camera, this algorithm also ensures the camera ends up at a suboptimal (if the optimal pose is unreachable) location among the locations already searched

## 2. Denoising Techniques Comparison

In this section, we will discuss more on various exiting image denoising techniques and conduct a comparison study of some techniques based on noise-reduction rate and high-frequency data loss.

Various denoising techniques have been proposed so far. A good noise removal algorithm ought to remove as much noise as possible while safeguarding edges. Gaussian white noise has been dealt with spatial filters, e.g., Gaussian filter, Mean filter and Wiener filter [43]. Image can be viewed as a matrix with variant intensity value at each pixel. The spatial filter is a square matrix that performs convolutional production with the Image matrix to produce filtered (noise-reduced) image, i.e.

$$g(i, j) = k * v(i, j) \qquad (1)$$

Where $v(i, j)$ is the original (noisy) image matrix entry at coordinate $(i, j)$. $k$ is the spatial filter (or kernel) and * accounts for convolution. $g(i, j)$ is the filtered (noisy-reduced) image matrix entry at coordinate $(i, j)$. All the spatial filters have the benefit of computationally efficient but shortcoming of blurring edges [25].

Also, image Gaussian noise reduction can be approached in wavelet domain. In conventional Fourier Transform, only frequency information is provided while the temporal information is lost in the transformation process. However, Wavelet transformation will keep both frequency and temporal information of the image [25]. The most widely used wavelet noise decreasing method is the wavelet threshold method, as it can not only get approximate optimal estimation of original image, but also calculate speedily and adapt widely [25]. The method can be divided into three steps. The first step is to decompose the noise polluted image by the wavelet transformation. Then wavelets of useful signal will be retained while the most wavelets of noise will be set zero according to the set threshold. The last step is to synthesize the new noise reduced image by inverse wavelet transformation of cleaned wavelets in the previous step [25,26]. The wavelet transformation method has benefits of keeping more useful details but more computationally complex than the spatial filters. Threshold affects the performance of the filter. Soft thresholding provides smoother results while hard threshold provides better edge preservation [25]. However, whichever threshold is selected, filters that operate in wavelet domain still filter out (or blur) some important high frequency useful details in original image, even though more edges are preserved than spatial filters.

All the aforementioned methods present ways to reduce noise in the image processing starting from a noisy image. We can approach this problem with the multiple noisy images taken from the same perspective. Assuming the same perspective ensures the same environmental conditions (illumination, temperature, etc.) that affect the image noise level. Given the same conditions, an image noise level taken at a particular time should be very similar to another image taken at a different time. This redundancy can be used for the purpose of improving image precision estimation in the presence of noise. The method that uses this redundancy to reduce noise is called signal averaging (or the image averaging in the application of the image processing) [27]. The image averaging has a natural advantage of retaining all the image details as well as reducing the unwanted noises, given that all the images for the averaging technique are taken from the same perspective.

The image averaging technique is illustrated in Figure 1. Assume a random, unbiased noise signal, and in addition, assume that this noise signal is completely uncorrelated with the image signal itself. As noisy images are averaged, the original true image is kept the same and the magnitude of the noise signal is compressed thus improving the signal-to-noise ratio. In Figure 1, we generated two random signals with the same standard deviation, and they are respectively represented by the blue and the red lines. The black line is the average of the two signals, whose magnitude is significantly decreased compared to each of the original signal. In general, we can come up with a mathematical relationship between the noise level reduction and the sample size for averaging. Assume we have $N$ numbers of Gaussian white noise samples with the standard deviation $\sigma$. Each sample is denoted as $z_i$, where $i$ represents $i^{th}$ sample signal. Therefore, we can acquire that:

$$var(z_i) = E(z_i{}^2) = \sigma^2 \qquad (2)$$

Where $var(\cdot)$ is the variance and $E(\cdot)$ is the expectation of the signal and $\sigma$ is the standard deviation.

By averaging $N$ Gaussian white noise signals, we can write:

$$var(z_{\text{avg}}) = var\left(\frac{1}{N}\sum_{i=1}^{N} z_i\right) = \frac{1}{N^2}N\sigma^2 = \frac{1}{N}\sigma^2 = \left(\frac{1}{\sqrt{N}}\sigma\right)^2 \qquad (3)$$

where $z_{\text{avg}}$ is the average of $N$ noise signals. Equation (3) demonstrates that a total number of $N$ samples are required to reduce the signal noise level by $\sqrt{N}$. Since our goal is to reduce the image noise within a fixed threshold (represented by a constant standard deviation), a smaller deviation in the original image requires much less samples to make an equivalent noise-reduction estimation. Thus, it is worthwhile for the camera to move around, rather than being stationary, to find the best locations where the image noise level estimation is small. In general, we can reduce the noise level as much as needed by taking more samples.

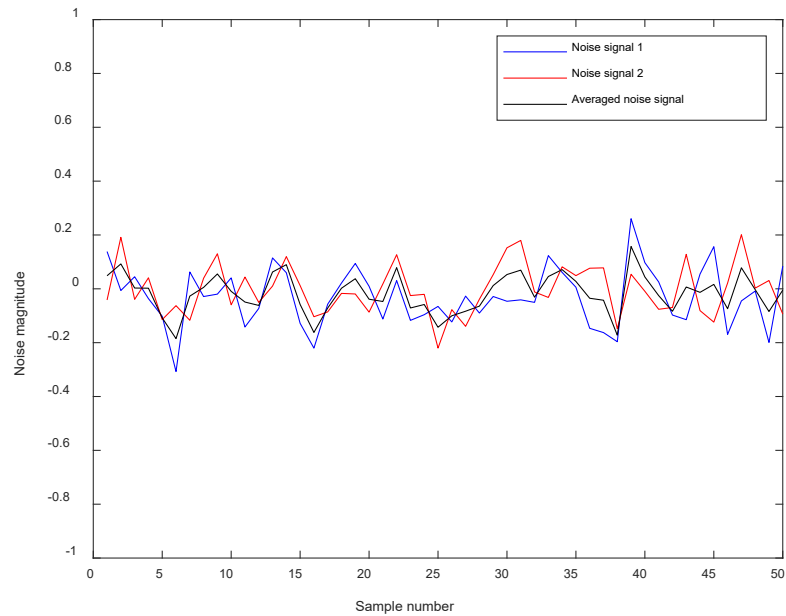

**Figure 1**. An example of the noise level reduction by image averaging.

A comparison of the above denoising methods is presented in the following. In figure 2, a default picture from MATLAB is polluted with a Gaussian distributed white noise with SNR equating to 5dB. A spatial filter (Gaussian filter), a wavelet filter (wavelet threshold method), and the image averaging technique (1000 average sample) have been applied to denoise the noisy image. Figure 2 shows all techniques succeed in reducing the noise to some extent while the wavelet filter and the image averaging technique perform better than the spatial filter regarding denoising. The difference between the wavelet filter and the image averaging technique is not obvious in figure 2 but becomes more significant in the frequency domain.

Figure 3 compares the performance in frequency domain of image averaging methods with different samples and figure 4 shows the difference in frequency domain after applying the above three denoising methods. The power intensity is separated from low to high as in those two figures. By comparing original image and noisy image, it can be observed that noise intensity is larger in higher frequency domain. Figure 3 shows that by averaging more samples, the noise level decreases in all frequencies. Especially, the averaged image is almost as same as the original one when 1000 samples are averaged.

Figure 4 shows the image generated from the Gaussian filter has lower power intensity than that of the original image in high frequency, which indicates that a lot of high frequency data in original image has been filtered out as well. Although it performs better than a Gaussian filter, the wavelet filter still gets rides of some details in the original image. Only the averaging method can reduce the noise level and, in addition, retains all information.

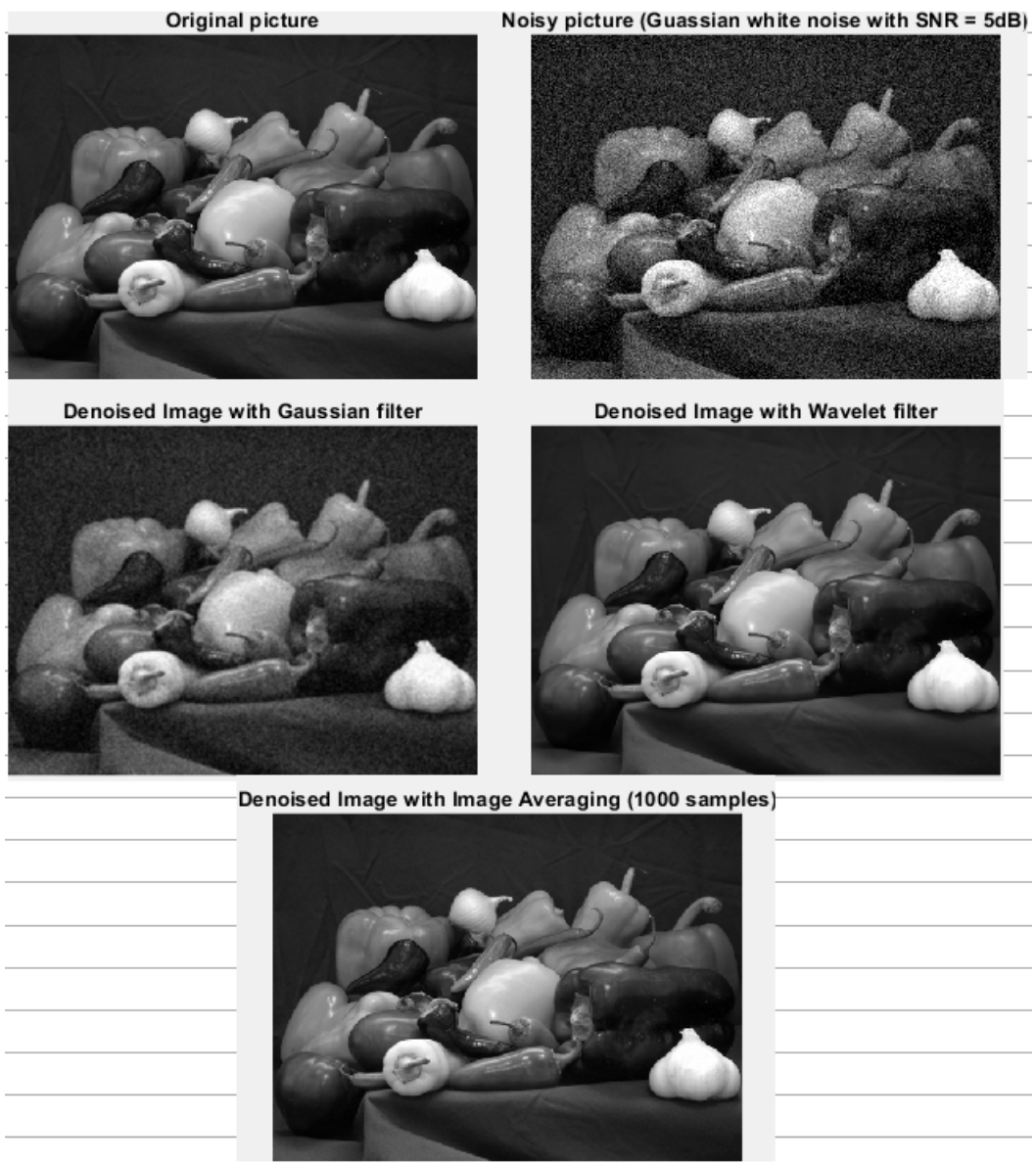

**Figure 2.** Original image, noisy image and denoised images

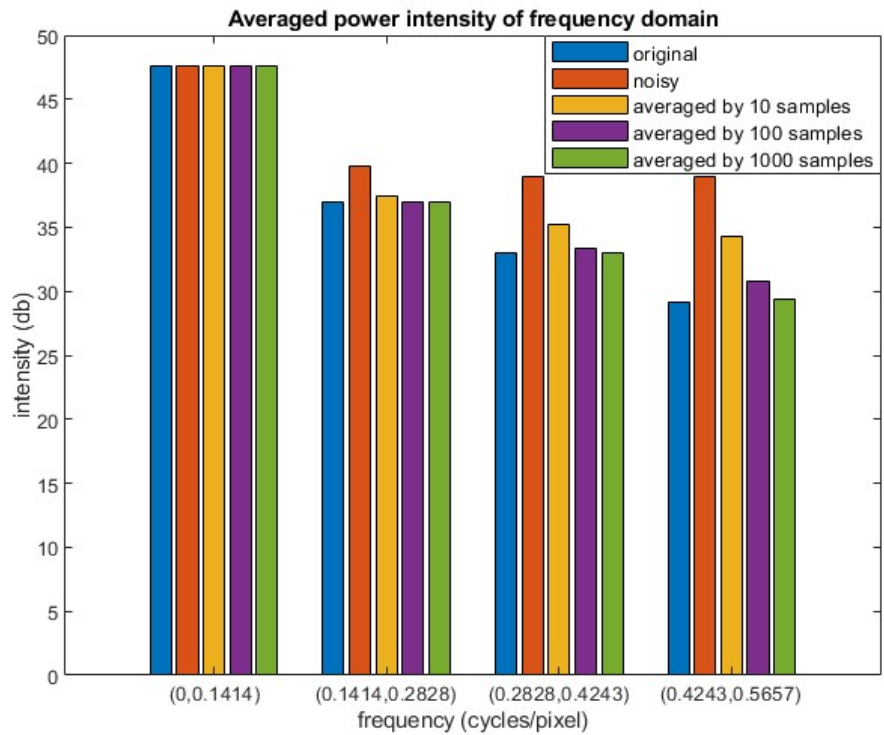

**Figure 3.** Frequency analysis of image averaging technique with varying samples

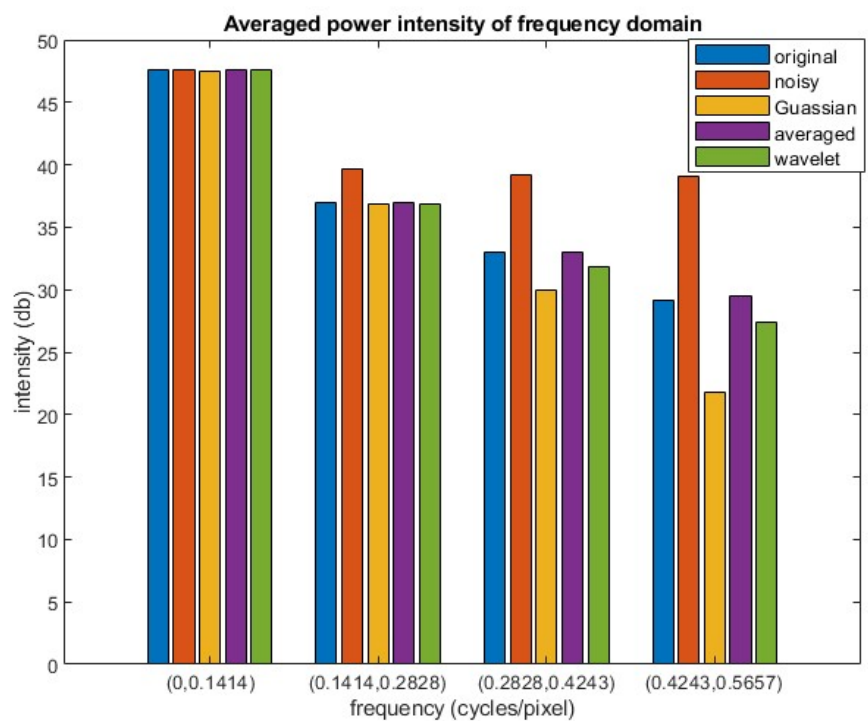

**Figure 4.** Frequency analysis of different denoising techniques

## 3. Algorithm Flowchart

The proposed algorithm is aimed to search and move the camera to the location where the least number of pictures are needed to reduce the noise level of the averaged image to a degree $\sigma_{noise\_reduced}$. A flowchart describing the algorithm is shown in Figure 5.

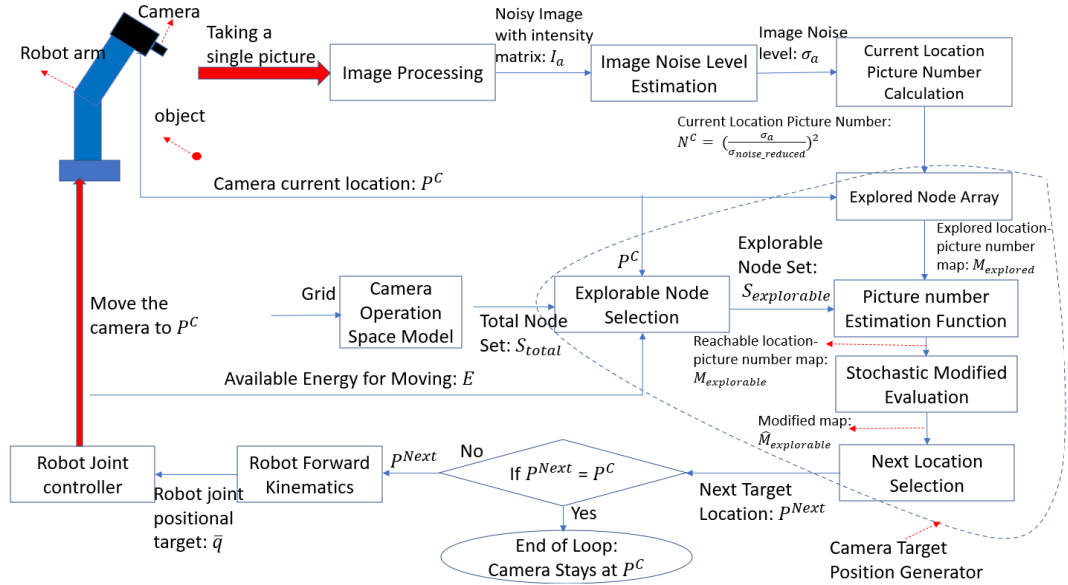

**Figure 5.** Algorithm flowchart.

A camera, mounted on a 6-joint robot manipulator, moves freely in 6 degrees of freedom (6 DoFs) in space. Assuming the camera is fixed in orientation towards the face of the object of interest, only the position (3 DoFs) of the camera can vary from the movement of the robot arm. The camera can only move to the locations within the maximum reach of the robot manipulator and the object can only be detectable within the view angle of the camera. Those constraints create a camera operational space (Figure 6a), which is then gridded with a certain resolution of distance. All the intersections of grids come up with a set of nodes ($S_{total}$), which are all possible candidates of locations that camera may move and search utilizing the algorithm. Each node is indexed and denoted as $Node_{index}$.

The algorithm iteratively commands the camera to take a picture at one location, generate the next target location, and move to that location until it finds the optimal location that requires the least number of pictures. In one iteration, the camera takes a photo at the current location $P^C$ and after image processing, which transforms a picture into pixel's intensity matrix, it produces a noisy image with an intensity matrix $I_a$. A previously developed algorithm [30] estimates the noise level $\sigma_a$ across the image. Then, at the current location of the camera, the number of pictures $N^C$ needed to reduce the noise level to $\sigma_{noise\_reduced}$ is calculated by the Equation: $N^C = (\frac{\sigma_a}{\sigma_{noise\_reduced}})^2$. Based on this information ($N^C$ at $P^C$), the algorithm then generates the next target location of the camera, $P^{Next}$. The decision-making process of generating $P^{Next}$ is illustrated in the dashed circle of the below figure. In general, this algorithm first determines, from current location $P^C$, the set of grid nodes that can be explorable ($S_{explored}$) in the next iteration based on available moving energy $E$. Then, a hash map regarding to the estimated number of pictures required for each explorable node ($M_{explorable}$) is evaluated from an estimation function that utilizes data from the hash map of all explored nodes ($M_{explored}$). A stochastic process further modifies the hash map $\hat{M}_{explorable}$ to add fidelity to the estimation model. In the end, the next node's location $P^{Next}$ is selected so that it has the smallest number of images required among all nodes in $\hat{M}_{explorable}$. If the next generated target position $P^{Next}$ is the same as the current position $P^C$, $P^C$ is the optimal location and the camera stays and observes the object at this location. Otherwise, the robot joint controller moves the robot joint angles to the targets $\bar{q}$ (calculated from forward kinematics [31] (Appendix Equation (B1) -(B5))) so that the camera will move to the next target position, $P^{Next}$.

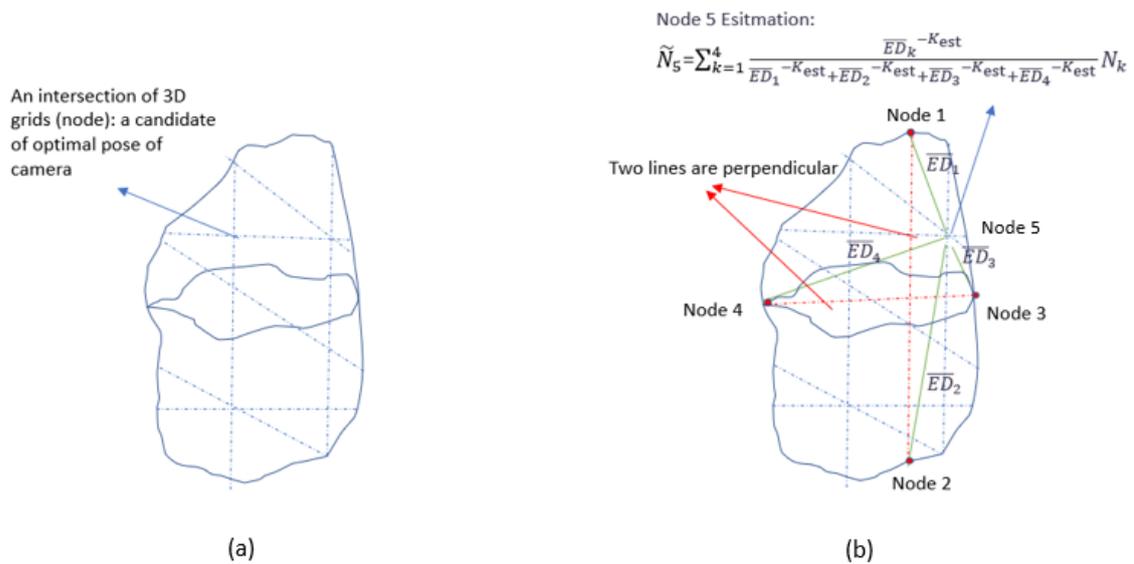

**Figure 6.** (a): The operational space of the camera is an arbitrary closed 3D geometry in the space shown in the plot. The gridded area is in 3D (blue dashed lines). (b): Four initial nodes are chosen so that their connection lines (red dashed line) are perpendicular. The picture number of Node 5 can be estimated by Equation (4) from four initial explored nodes. $\overline{ED}$ represents Euclidean distance (green line).

## 4. Core Algorithm

The core of this algorithm is to generate the target locations to be explored in the camera's operational space. The details of this algorithm are shown in the area circled with dashed lines in Figure 5. In this section, we will put efforts into the development of the picture number estimation function, the explorable nodes set, the stochastic modified evaluation function, the next location selection process, and the estimated energy cost function.

### 4.1. Picture Number Estimation Function

In each iteration of the algorithm, the camera moves to a new node with location $P^C$, and the algorithm evaluates the number of pictures $N^C$ needed at this location. The node ($Node_{index}$) is then set to be explored and saved in a set $S_{explored}$. Each node that has been explored in iterations is one-on-one paired with the values: location, and picture numbers of that node. All pairs that are expressed as $Node_{index}^{explored}:(P_{index}^{explored}, N_{index}^{explored})$ are saved in a hash map $M_{explored}$. Then $M_{explored}$ is used to develop an estimation function that can estimate the number of pictures needed for the rest unexplored nodes. The estimated picture number calculated for all nodes in space is denoted as $S_{\tilde{N}}$. Expressions of any element in $S_{\tilde{N}}$ are:

*For any node index $x$ in the Total Node Set: $S_{total} = \{Node_x | x \in (1, 2, \dots m)\}$, where $m$ is the total number of nodes.*

$$\tilde{N}_x = \begin{cases} \sum_{i \in S_{index-exp}} \dfrac{(\overline{ED}(P_i^{exp}, P_x)^\wedge(-K_{est}))}{\sum_{j \in S_{index-exp}}((\overline{ED}(P_j^{exp}, P_x)^\wedge(-K_{est}))} N_i^{explored} & , if\ Node_x \notin S_{explored} \\ N_x^{explored}, \quad if\ Node_x \in S_{explored} \end{cases} \tag{4}$$

Then: $S_{\tilde{N}} = \{\tilde{N}_x | x \in (1,2,\dots m)\}$ (5)

where $K_{est}$ is a positive constant parameter

$$S_{index-exp} = \{index | Node_{index} \in S_{explored}\}, \tag{6}$$

$$\overline{ED}(P_a, P_b) = \sqrt{(P_{ax} - P_{bx})^2 + (P_{ay} - P_{by})^2 + (P_{az} - P_{bz})^2},$$

with $P_a, P_b$ are 3D locations of $Node_a\ and\ Node_b$: (7)

$$P_a = (P_{ax}, P_{ay}, P_{az})\ and\ P_b = (P_{bx}, P_{by}, P_{bz})$$

Equation (4) is a weighted function concerning each explored node. If one explored node $Node_i^{explored}$ is closer to the node $Node_x^{explored}$, its picture number $N_i^{explored}$ has a larger weight (effect) on the estimation of $\tilde{N}_x$. Thus, a minus sign of $K_{est}$ is utilized to indicate the negative correlation between the Euclidean distance and weight.

Equation (4) shows that at least some initial nodes are required to be explored before their values can be used to estimate the picture numbers in other nodes. For the estimation function to work properly by covering all nodes in the camera operational space, four initial nodes are selected in the following steps:

1. Find the first two nodes on the boundary of operational space so that the Euclidean distance between these two nodes is the largest.
2. Select other two nodes whose connection line is perpendicular to the previous connection line and the Euclidean distance between these two nodes is also the largest among all possible node pairs.
3. Move the camera to those four nodes in space with proper orientation. Take a single image at each location and estimate the number of pictures required at those locations. And save all four nodes in $S_{explored}$ and their values in $M_{explored}$.

Figure 6b shows the selection of four initial nodes in a camera operational space and shows an example of the use of the estimation function to estimate the picture number in one node.

## 4.2. Explorable Node Selection

A stage in the algorithm, Explorable Node Selection, generates a set of nodes $S_{explorable}$, which contains all the camera's next explorable locations from the current location $P^c$ with available current energy for moving $E_{bound}$. The explorable set is found in the following steps (Figure 7):

1. With current energy bound $E_{bound}$, find all nodes in the space that can be reached from the current node $P^c$.

In other words, all feasible nodes are in the set:

$$S_{feasible} = \{Node_{index} | E(P^c, P_{index}) \le E_{bound}\} \tag{8}$$

where $P$ is the position of a node and $E(P_a, P_b)$ is the estimated energy cost from $P_a$ to $P_b$.

When the current energy bound $E_{bound}$ is greater than a predefined energy threshold $E_T$ ($E_{bound} > E_T$), then the algorithm is safe to explore all nodes in the feasible range $S_{feasible}$ in the next iteration loop. In other words:

$$S_{explorable} = S_{feasible}, \qquad if\ E_{bound} > E_T \tag{9}$$

*Note*: $E(P_a, P_b)$ is generated from a dynamic model of a robot arm control system. The details of this function are developed in the following section of this paper.

2. When $E_{bound} \leq E_T$, step 2 and step 3 are applied in finding the explorable set. In the feasible set, look at all explored nodes and find the one that has the smallest number of pictures. In other words:

$$S_{feas\&exp} = \{Node_{index} | Node_{index} \in S_{feasible} \cap S_{explored}\} \qquad (10)$$

$$\exp\_min = \{index | N_{\exp\_min} = \min(N_{index} | Node_{index} \in S_{feas\&exp})\} \qquad (11)$$

where $S_{explored}$ is the set of all explored nodes.

$S_{feas\&exp}$ is the intersection set between $S_{feasible}$ and $S_{explored}$.

3. The estimation function may not give accurate results for some unexplored nodes. Therefore, it is possible that the algorithm may make the camera end up in a node that has a large number of pictures in some iteration loops. Because of that, our algorithm needs to make sure at worst it has enough energy to go back to the best (minimum number of pictures) node that has been explored when the available energy is low ($E_{bound} \leq E_T$). This further reduces the feasible set $S_{feasible}$. The explorable set $S_{explorable}$ can be written as:

$$S_{explorable} = \{Node_{index} | E_1 + E_2 \leq E_{bound}\}, \qquad if\ E_{bound} \leq E_T \qquad (12)$$

$$E_1 = E(P^c, P_{index}),\ E_2 = E(P_{index}, P_{\exp\_min}) \qquad (13)$$

Figure 7 illustrates the selection of explorable nodes set.

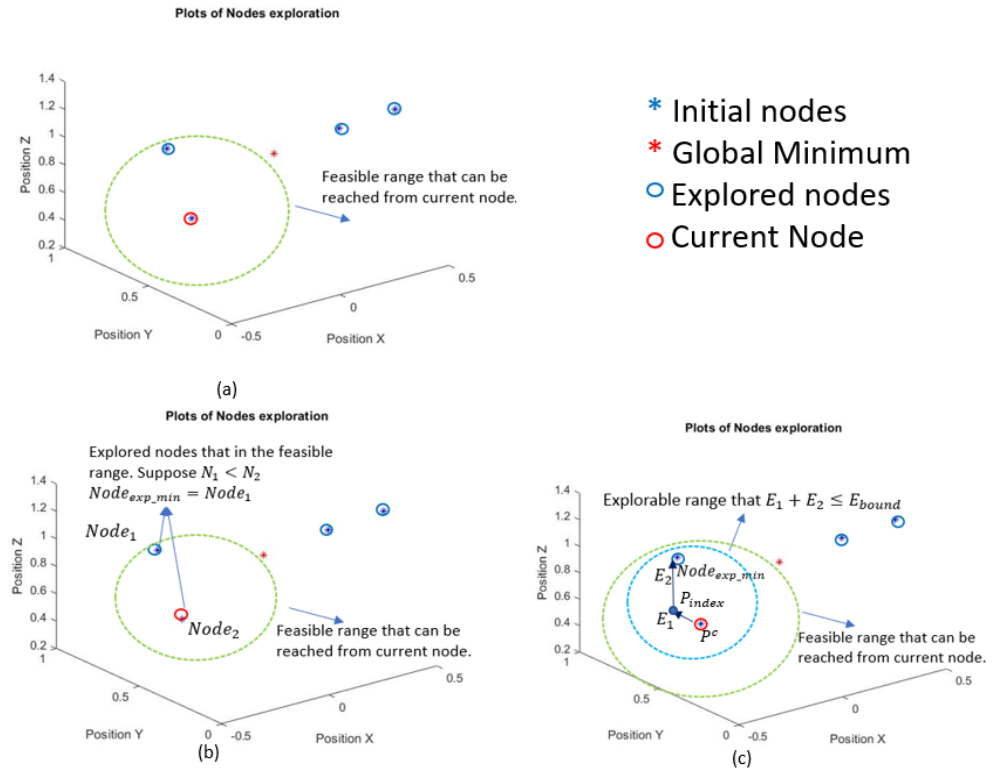

**Figure 7.** Find the explorable set $S_{explorable}$ when the global minimum (red star) is inaccessible. (a). Step 1: Obtain the feasible range $S_{feasible}$ (green dashed line). (b). Step 2: Find the explored node with a minimum picture number in the feasible range. Initial nodes (blue stars) are explored (blue circles). $Node_1$ is $Node_{exp\_min}$.(c). Step 3: Obtain explorable range when $E_{bound} \leq E_T$. The explorable range $S_{explorable}$ (blue dashed line) is a subset of $S_{feasible}$.

### 4.3. Stochastic Modified Evaluation

For all nodes in $S_{explorable}$, we set up a hash map $M_{explorable}$. For each node, its location and number of required pictures for that node are given as: $Node_{index}^{explorable}$: $(P_{index}^{explorable}, \widetilde{N}_{index}^{explorable})$. The number of pictures required for nodes in $S_{explorable}$ are calculated from the estimation function in Equation (4).

The algorithm tends to select nodes located around the explored node that has the minimum number of pictures because, from Equation (4), those nodes tend to have the smallest estimated number of pictures. However, this process does not guarantee reaching the global minimum, therefore, minimizing the number of pictures. To sufficiently explore unknown nodes as well as exploit information from already explored nodes, a stochastic process is introduced to modify the picture number estimation function given in Equation (4).

Assume when the locations of a camera in space deviate with a smaller amount from each other, the difference of real picture number values at those locations is also smaller. Thus, among all unexplored nodes, the estimated picture number of an unexplored node, which is closer to the explored node, is more deterministic. To emphasize this feature, we make the estimated picture number follow a normal distribution with a mean $\mu_{\widetilde{N}}$ being equal to the value from Equation (4) and a standard deviation $\sigma_{\widetilde{N}}$. As an unexplored node is further away from explored nodes, the larger value of $\sigma_{\widetilde{N}}$ should be assumed, which means more indeterministic estimation (illustrated in Figure 8). The following Equations (14)-(16) summarize the above discussion:

For any nodes $Node_{index=x}^{explorable}$ in $S_{explorable}$, its picture number estimation should follow the following normal distributions:

$$P(Z_x) = \mathcal{N}(\mu_{\widetilde{N}_x}, \sigma_{\widetilde{N}_x}) \tag{14}$$

$$\mu_{\widetilde{N}_x} = \widetilde{N}_x^{explorable} \tag{15}$$

$$\sigma_{\widetilde{N}_x} = K_{sd} * \overline{ED}_{min} \tag{16}$$

where $P(Z_x)$ is the probability of a random variable $Z$ at $Node_x^{explorable}$. $\widetilde{N}_x^{explorable}$ is the estimated value from Equation (4) of $Node_x^{explorable}$. $K_{sd}$ is a constant parameter, and $\overline{ED}_{min}$ is the smallest Euclidean distance among distances between that node and explored nodes.

As discussed above, the algorithm ensures that the camera at any iteration, always has enough energy to move back to the previous best node $N_{exp\_min}$. Therefore, a new distribution can be generated, which sets values bigger than the minimum ($N_{exp\_min}$) in the original distribution to be the minimum value. Then Equations (14) to (16) are modified with a new random variable $Z_x^{new}$ and its expectations as follows:

$$Z_x^{new} = \begin{cases} Z_x, & if \ Z_x < N_{exp\_min} \\ N_{exp\_min}, & if \ Z_x \geq N_{exp\_min} \end{cases} \tag{17}$$

$$P(Z_x^{new}) = P(Z_x) = \mathcal{N}(\mu_{\widetilde{N}_x}, \sigma_{\widetilde{N}_x}) \tag{18}$$

$$\widehat{N}_x^{explorable} = \sum [Z_x^{new} * P(Z_x^{new})] \tag{19}$$

where $\widehat{N}_x^{explorable}$ is the stochastic modified estimated value for the node $Node_x^{explorable}$.

Then a new hash map $\widehat{M}_{explorable}$ is set up with pairs as $Node_{index}^{explorable}$: $(P_{index}^{explorable}, \widehat{N}_{index}^{explorable})$. The next target node to be explored is the one that has the smallest modified estimated value in the map:

$$index_{Next} = \{index | \widehat{N}_{index_{Next}}^{explorable} = \min(\widehat{N}_{index}^{explorable})\} \tag{20}$$

$$P^{Next} = P_{index_{Next}}^{explorable} \tag{21}$$

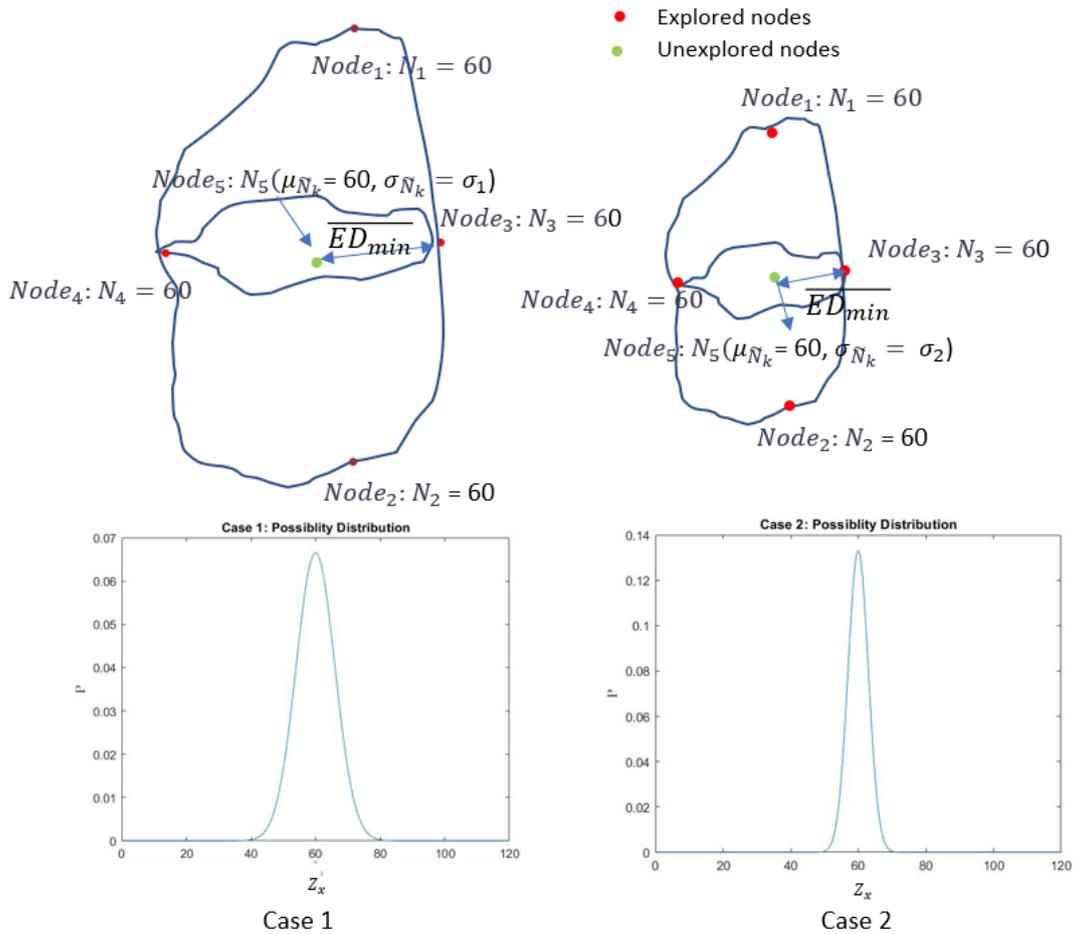

Figure 8. Both case1 and case2 have the same estimated value of an unexplored node (green point) from Equation (4). However, the node in case 1 has a larger distance to explored nodes (red points) compared to that in case 2. Therefore, we set different standard deviations ($\sigma_1 > \sigma_2$) to differentiate possibility distributions. This approach distinguishes nodes that have the same estimations from Equation (4) to have different likelihood to be explored in the next iteration.

### 4.4. Estimated Energy Cost Function

As discussed above, an estimated energy cost function $E(P_a, P_b)$ calculates the estimated energy cost from $P_a$ to $P_b$. This function is used to find feasible node set $S_{feasible}$ and explorable node set $S_{explorable}$. In this section, the equations for the cost function are derived from the robot positioning-controlled system's response. Also, trade-offs between energy cost and settling time of moving are discussed.

The energy of moving the camera between two locations from $P_{initial}$ to $P_{final}$, results from the energy cost of DC motors in 6 DOFs robot arm's each joint rotating from joint angles $q_{initial}$ to $q_{final}$. Therefore, $E(P_{initial}, P_{final})$ can be described as the sum of the time integration of rotational power in each joint. That is:

$$E(P_{initial}, P_{final}) = \sum_{i=1}^{6} \int_{t_0^i}^{t_f^i} V^i \cdot I^i \cdot dt \tag{22}$$

where $V^i$ is the voltage and $I^i$ is the current in the DC motor's circuit of the $i^{th}$ joint. $t_0^i$ and $t_f^i$ are initial time and final time of the $i^{th}$ joint moving from its initial angle $q_{initial}^i$ to its final angle $q_{final}^i$.

Without detailed derivation, the dynamic model of a 6dofs revolutionary robotic manipulator and DC motors can be expressed as [31]:

$$\frac{1}{r_k}J_{m_k}\ddot{q}_k + \sum_{j=1}^{6}d_{j,k}\ddot{q}_j + \sum_{i,j}^{6}c_{i,j,k}(q)\dot{q}_i\dot{q}_j + \frac{1}{r_k}B\dot{q}_k + \Phi_k(q) = \frac{K_m}{r_kR}V^i \tag{23}$$

$$J_m = J_a + J_g, \qquad B = B_m + \frac{K_bK_m}{R} \tag{24}$$

$$c_{i,j,k} = \frac{1}{2}\left(\frac{\delta d_{k,j}}{\delta q_i} + \frac{\delta d_{k,i}}{\delta q_j} - \frac{\delta d_{i,j}}{\delta q_k}\right), \ \Phi_k = \frac{\delta V}{\delta q_k} \tag{25}$$

$$i,j,k \in (1,2,3,4,5,6) \tag{26}$$

where Equations (23) and (24) express the $k^{th}$ joint dynamics equation. $q_i/q_j/q_k$ is the joint revolute variable. $J_a$ $and$ $J_g$ represent the moment of inertia of the motor, and the gear of the model. $r_k$ is gear ratio at $k^{th}$ joint, and $B_m$ is the damping effect of the gear. $d_{i,j}$ represents the entry of inertial matrix of the robot manipulator at $i^{th}$ row and $j^{th}$ column. $c_{i,j,k}$ is Christoffel symbols and for a fixed $k$, with $c_{i,j,k} = c_{j,i,k}$. And $\Phi_k$ is the derivative of potential energy $V$ with respective to $k^{th}$ joint variance. $K_m$ is the torque constant in $N-m/amp$, $K_b$ is the back emf constant, and $R$ is Armature Resistance.

Take $u^i$ as the actuator input to the dynamic system (23) measured at $i^{th}$ joint from a designed controller. Therefore, $u^i$ equals to the right-side of Equation (23). That is:

$$V^i = \frac{rR}{K_m}u^i \tag{27}$$

In addition, without derivation, the expression of current in the Laplace domain [31] is:

$$(Ls + R)I^i(s) = V^i(s) - \frac{K_b}{r}sq^i(s) \tag{28}$$

Then: $I^i(s) = \frac{1}{(Ls+R)}V^i(s) - \frac{K_b}{r}\frac{s}{(Ls+R)}q^i(s) \tag{29}$

Take the inverse Laplace: $I^i(t) = \frac{1}{L}e^{-\frac{R}{L}t} * V^i(t) - \frac{K_b}{r}(\frac{1}{L} - \frac{R}{L^2}e^{-\frac{R}{L}t}) * q^i(t) \tag{30}$ where $L$ is armature inductance, $K_b$ is back emf constant, and $*$ is the convolutional multiplication. Therefore, the instant current in the time domain $I^i(t)$ is a function of the instant voltage $V^i(t)$ and the instant angle q$(t)$.

Various controller designs, such as PID controller [32], and Youla controller [33], of six DoFs revolute robotic manipulators have been well developed in many papers. In this paper, we use a previous Youla controller design [33] with feedback linearization (Figure 9) as the position controller of the robot manipulator that holds the camera.

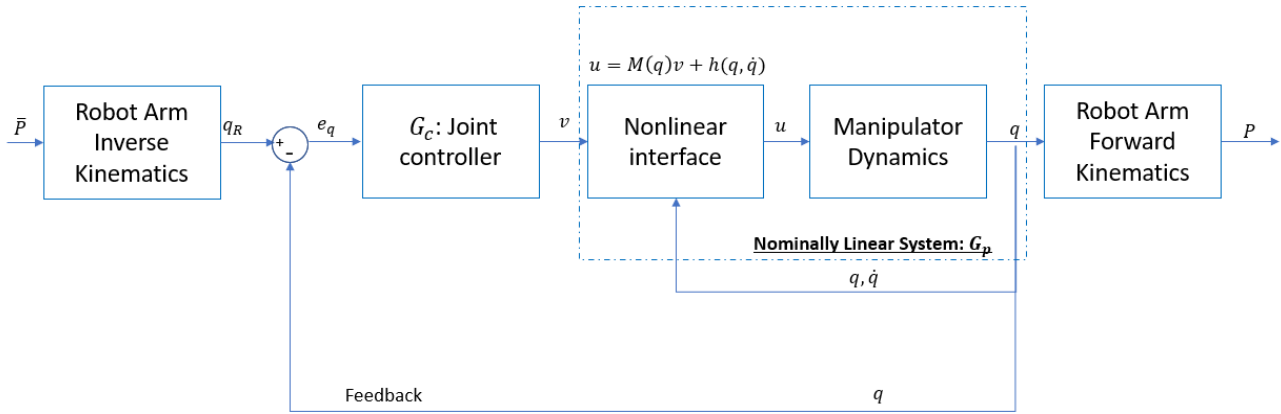

**Figure 9.** The block Diagram of feedback linearization Youla control design used for the joint control loop.

From Figure 5, the actuator input $u^i$ can be derived from a virtual input $v^i$ so that:

$$u^i = M(q^i(t))v^i + H(q^i(t), \dot{q}^i(t)) \tag{31}$$

where M, and H are nonlinear functions of $q^i(t)$, and $\dot{q}^i(t)$, the first derivative of $q^i(t)$.

Without showing the controller development in this paper, the 1 DoF controller transfer function $G_c(s)$ and 1 DoF nominal plant $G_p(s)$ with $v^i$ as the virtual actuator input in the feedback linearization are expressed as follows:

$$G_c(s) = \frac{3\tau_{in}^2 s + 1}{\tau_{in}^3 s + 3\tau_{in}^2} \tag{32}$$

$$G_p(s) = \frac{1}{s^2} \tag{33}$$

where $\tau_{in}$ is a constant parameter in the controller design.

Then the following transfer functions can be calculated as:

$$\frac{v^i(s)}{q^i_{final}(s)} = \frac{G_c}{1 + G_c G_p} \tag{34}$$

$$\frac{q^i(s)}{q^i_{final}(s)} = \frac{G_c G_p}{1 + G_c G_p} \tag{35}$$

$$\frac{\dot{q}^i(s)}{q^i_{final}(s)} = \frac{s G_c G_p}{1 + G_c G_p} \tag{36}$$

By taking inverse Laplace transform, and $q^i_{final}$ as a step input, the following Equations (37)-(39) in the time domain are:

$$v^i(t) = (q^i_{final} - q^i_{intial})(\frac{1}{\tau_{in}^4}t^2 e^{-\frac{t}{\tau_{in}}} + \frac{-5}{\tau_{in}^3}t e^{-\frac{t}{\tau_{in}}} + \frac{3}{\tau_{in}^2}e^{-\frac{t}{\tau_{in}}}) \tag{37}$$

$$q^i(t) = (q^i_{final} - q^i_{intial})(\frac{1}{\tau_{in}^2}t^2 e^{-\frac{t}{\tau_{in}}} - \frac{1}{\tau_{in}}t e^{-\frac{t}{\tau_{in}}} - e^{-\frac{t}{\tau_{in}}} + 1) + q^i_{intial} \tag{38}$$

$$\dot{q}^i(t) = (q^i_{final} - q^i_{intial})(\frac{3}{\tau_{in}^2}t e^{-\frac{t}{\tau_{in}}} - \frac{1}{\tau_{in}^3}t^2 e^{-\frac{t}{\tau_{in}}}) \tag{39}$$

With expressions of $v^i(t), q^i(t)$, and $\dot{q}^i(t)$, Equations (27) and (31) shows that:

$$V^i(t) = F(q^i_{intial}, \; q^i_{final}, t) \tag{40}$$

and combine Equations (30) and (40):

$$I^i(t) = G(q^i_{intial}, \; q^i_{final}, t) \tag{41}$$

where F, and G are nonlinear functions of $q^i_{initial}$, $q^i_{final}$, and $t$.

It has been shown so far that the estimated energy cost $E(P_{initial}, P_{final})$ from location $P_{initial}$ to $P_{final}$ is a function of $q_{initial}$ and $q_{final}$.

$q_{initial}$ and $q_{final}$ are derived from the inverse kinematics process [31] (Shown in Appendix Equations (B6) -(B13)), that is:

$$q_{initial} = inverkinematics(P_{initial}) \tag{42}$$

$$q_{final} = inverkinematics(P_{final}) \tag{43}$$

And $q_{initial} = \left[q^i_{initial} \middle| i \in (1,2,3,4,5,6)\right]$, $q_{final} = \left[q^i_{final} \middle| i \in (1,2,3,4,5,6)\right]$ (44)

Development of Equations (37) to (39) assumes that the target angle of each joint $q^i_{final}$ is a step input. A more realistic assumption is to set $q^i_{final}$ as a delayed input.

$$q^i_{final}(t) = (1 - e^{-\frac{t}{\tau_{delay}}})q^i_{final} \tag{45}$$

where $\tau_{delay}$ is a time constant that measures the time delay of the target in the real positioning control. $\tau_{delay} \geq 0$, and when $\tau_{delay} = 0$, it indicates no delay exists in the input.

The Laplace form of (45) is:

$$q^i_{final}(s) = (\frac{1}{s} - \frac{1}{s + \frac{1}{\tau_{delay}}})q^i_{final} \tag{46}$$

With new expression of $q^i_{final}(s)$, Equation (37) to (39) can be developed as:

$$v^i(t) = (q^i_{final} - q^i_{initial})(\frac{A_v}{\tau_{in}^3}t^2 e^{-\frac{t}{\tau_{in}}} + B_v\frac{B_v}{\tau_{in}^2}te^{-\frac{t}{\tau_{in}}} + \frac{C_v}{\tau_{in}}e^{-\frac{t}{\tau_{in}}} + \frac{D_v}{\tau_{delay}}e^{-\frac{t}{\tau_{delay}}}) \tag{47}$$

$$A_v = \frac{1}{\tau_{in}} - \frac{-\tau_{delay}}{\tau_{in}(\tau_{in}-\tau_{delay})} \qquad B_v = \frac{-5}{\tau_{in}} - \frac{7\tau_{in}\tau_{delay}-5\tau_{delay}^2}{\tau_{in}(\tau_{in}-\tau_{delay})^2}$$

$$C_v = \frac{3}{\tau_{in}} - \frac{\tau_{delay}(-8\tau_{in}^2+9\tau_{in}\tau_{delay}-3\tau_{delay}^2)}{\tau_{in}(\tau_{in}-\tau_{delay})^3} \qquad D_v = -\frac{3\tau_{in}\tau_{delay}-\tau_{delay}^2}{(\tau_{in}-\tau_{delay})^3}$$

$$q^i(t) = (q^i_{final} - q^i_{initial})(\frac{A_Q}{\tau_{in}^3}t^2 e^{-\frac{t}{\tau_{in}}} + \frac{B_Q}{\tau_{in}^2}te^{-\frac{t}{\tau_{in}}} + \frac{C_Q}{\tau_{in}}e^{-\frac{t}{\tau_{in}}} + \frac{D_Q}{\tau_{delay}}e^{-\frac{t}{\tau_{delay}}}+1) + q^i_{initial} \tag{48}$$

$$A_Q = \tau_{in} + \frac{\tau_{in}\tau_{delay}}{(\tau_{in}-\tau_{delay})} \qquad B_Q = -\tau_{in} - \frac{3\tau_{in}^2\tau_{delay}-\tau_{in}\tau_{delay}^2}{(\tau_{in}-\tau_{delay})^2}$$

$$C_Q = -\tau_{in} - \frac{-3\tau_{in}^2\tau_{delay}^2+\tau_{in}\tau_{delay}^3}{(\tau_{in}-\tau_{delay})^3} \qquad D_Q = -\frac{3\tau_{in}\tau_{delay}^3-\tau_{delay}^4}{(\tau_{in}-\tau_{delay})^3}$$

$$\dot{q}^i(t) = (q^i_{final} - q^i_{initial})(\frac{A_{\dot{Q}}}{\tau_{in}^4}t^2 e^{-\frac{t}{\tau_{in}}} + \frac{B_{\dot{Q}}}{\tau_{in}^3}te^{-\frac{t}{\tau_{in}}} + \frac{C_{\dot{Q}}}{\tau_{in}^2}e^{-\frac{t}{\tau_{in}}} + \frac{D_{\dot{Q}}}{\tau_{delay}^2}e^{-\frac{t}{\tau_{delay}}}) \tag{49}$$

$$A_{\dot{Q}} = -A_Q \qquad B_{\dot{Q}} = 2A_Q - B_Q$$

$$C_{\dot{Q}} = B_Q - C_Q \qquad D_{\dot{Q}} = -D_Q$$

A simulation scenario is set up to calculate how estimated energy cost changes with varying $\tau_{delay}$. Set $P_{initial} = [-0.30, 0.05, 1.20]$, and $P_{final} = [-0.45, 0.45, 1.20]$ and use the Equations (42)-(44), (47)-(49). Figure 10 shows the response of one angle $q^1(t)$ with varying $\tau_{delay}$ and Table 1 presents the estimated energy cost: $E(P_{initial}, P_{final})$ and settling time: $t_s$ with varying $\tau_{delay}$.

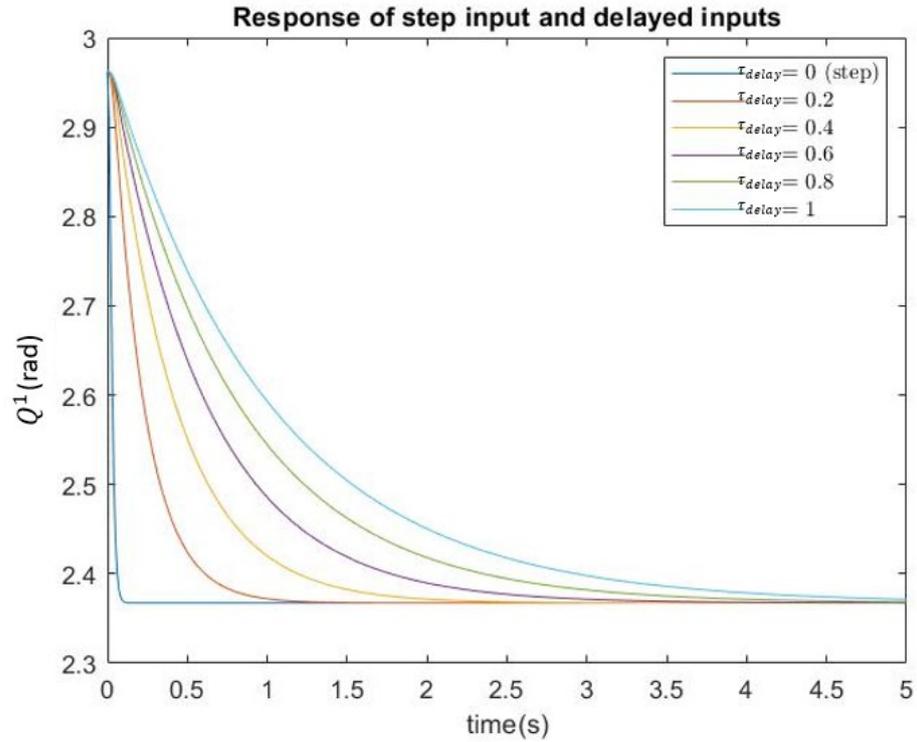

**Figure 10.** Response of $q^1(t)$ with varying $\tau_{delay}$.

**Table 1.** Estimated Energy Cost and Settling Time with Varying Delay Constant.

| $\tau_{delay}$ [s] | $E(P_{initial}, P_{final})$ [ws] | $t_s$ [s] |
|:---:|:---:|:---:|
| 0 | 10.895 | 0.07 |
| 0.2 | 0.102 | 0.73 |
| 0.4 | 0.030 | 1.42 |
| 0.6 | 0.016 | 2.11 |
| 0.8 | 0.011 | 2.80 |
| 1 | 0.009 | 3.49 |

Note: $E(P_{initial}, P_{final})$ is measured as [watts seconds]

Table 1 shows a tradeoff between the estimated energy cost and settling time; reducing the energy cost of moving the camera inevitably increases the response time. This finding matches the results in Figure 6.

Systematic delays are inevitable in controlled systems' design. Delays are incurred in many sources such as time required for sensors to detect and process changes, for actuators response to control signals, and for controllers to process and compute signals. In a real manufacturing environment, larger delays cause slower motion of the camera when searching the area but slower response results in less energy consumption from actuators in the manipulator.

If the delay of input is given or can be measured, the estimated energy cost can be calculated through the process in this section. However, if the value is unknown, the delay constant must be chosen and decided based on the following criteria:

- Select small $\tau_{delay}$ for a conservative algorithm that searches a small area but ensures it ends up with the minimum that has been explored.

- Select large $\tau_{delay}$ for an aggressive algorithm that searches a large area but risks not ending up at the minimum that has been explored.

In this section, the estimated energy cost function is well developed. However, the estimation function is developed from an ideal camera movement. The accuracy of the estimation can be negatively influenced by some unmodeled uncertainties such as backlash from gears in robot manipulators, unmodeled compliance components from joint vibration, etc. To tackle the problem, we can derive an online updated model of the estimation function by comparing estimated voltage and current and real-time measurements. The parameter values of motors and gears used in the simulations are summarized in Appendix Table A4.

## 5. Simulation Setup

The developed algorithm has been simulated on an application in automated manufacturing (Figure 11). This is a multi-robot system composed of a visual system and a tool manipulation system. In the visual system, a camera is mounted on an elbow robot arm while a tool is held by the end-effector of the robot manipulator arm. The goal of the visual system is to provide a precise estimation of the tool pose so that the tool manipulator can control the pose with guidance from the visual system. The first step of the manufacturing process is to move the camera in the space and search for the best location so that the number of images required for averaging is minimized to reduce the image noises within an acceptable limit. Then the camera is fixed and provides continuous vision data to the tool manipulator. Therefore, the control and operation of two manipulators are asynchronous. The model of both robot manipulators is ABB IRB 4600 [34] and the model of the stereo camera we choose in this project is Zed 2 [35].

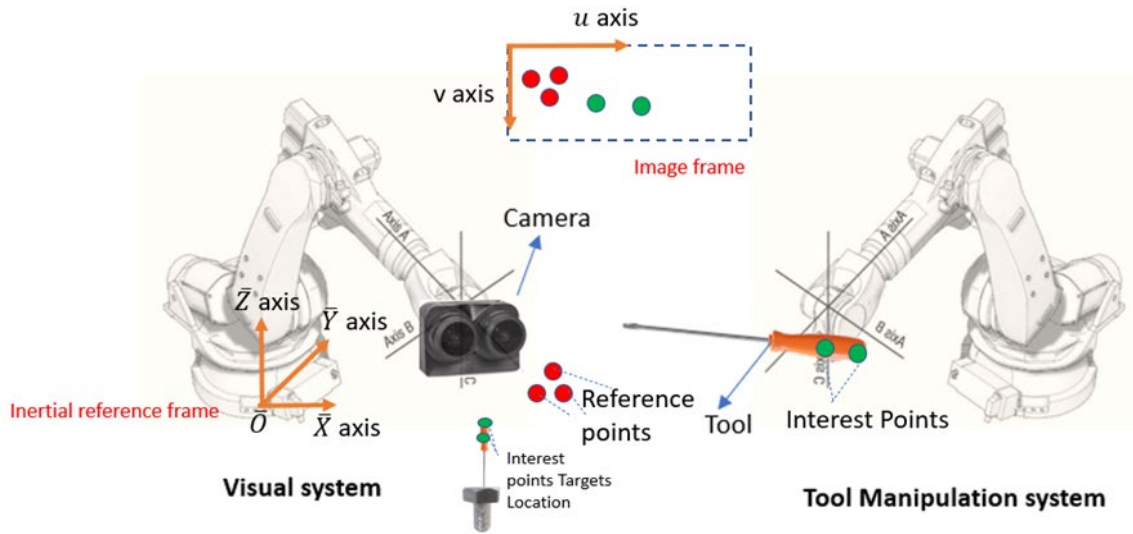

**Figure 11.** The topology of the multi-robotic system for accurate positioning control.

We first define the camera's operational space in this application and then provide the simulation results of a made-up scenario in the next.

In the above discussion, a camera operational space is gridded with nodes and each node is a potential candidate of the camera's optimal position. The camera operational space is defined by three geometric constraints below:

1. The camera can only be allowed to locations where fiducial markers that are attached to the tool are recognizable on the image frame. Therefore, the fiducial markers are in the angle of view of the camera, and the distance between the markers and the camera center is within a threshold.

2. The camera can only be allowed to locations within the reach of the robot arm.

3. The camera can only be allowed in locations where the visual system and the tool manipulation system do not physically interfere with each other.

From the specifications of the stereo camera and the robot arm (Appendix Table A1 and A3), the geometry and dimensions of the camera operational space are analyzed in the following part of the subsections.

## 5.1. Reachable and Dexterous Workspace of Two-Hybrid Systems

In the spherical wrist model of the robot arm, three rotational axes represent the pitch, roll, and yaw of the end-effector independently. As shown in Figure 12, those three axes intersect at point $H$. Point $P$ is the location of the end-effector and it's the place where the camera or tool is mounted. For the elbow manipulator mounted with the stereo camera, the camera is placed so that its optical center line coincides with axis C. The location of camera optical center C is determined and a good estimation of it can be found in this paper [36]. For now, assume the optical center is in the middle of $\overline{PJ}$ (set $\overline{PC} = 17\ mm$). For the elbow manipulator mounted with the tool, the screwdriver is attached at the point $P$. Point J indicates the far end of the object (tool or camera) attached to the end effector. Point H is kept stationary no matter which axis rotates. Therefore, the task of positioning and rotation is decoupled in the spherical wrist robot model.

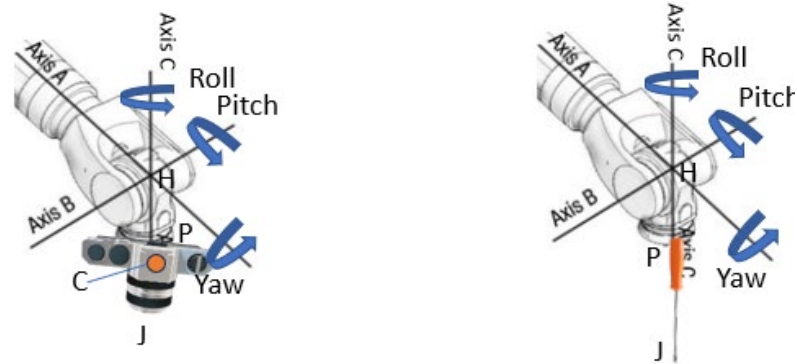

**Figure 12.** Spherical Wrist Model.

The idea of reachable and dexterous workspaces of an ideal elbow manipulator has been introduced in the paper [37]. An ideal elbow manipulator is a manipulator whose angles of rotation are free to move in the whole operational range $[0, 2\pi]$. However, a realistic elbow manipulator is limited to moving its joints within certain ranges of angle.

The workspace of an ideal elbow manipulator (Figure 13) is a sphere centered at the joint 1 of the manipulator, denoted as point $o$. The reachable workspace concerning the center o of a manipulator is the aggregate of all possible locations of the point $J$ attached with the end-effector and is denoted as $W_o(J)$. The dexterous workspace to the center o of a manipulator is the aggregate of all possible locations that point $J$ can reach all possible orientations of the end-effector and is denoted as $W_o^d(J)$. The reachable workspace of $H$ is denoted as $W_o(H)$.

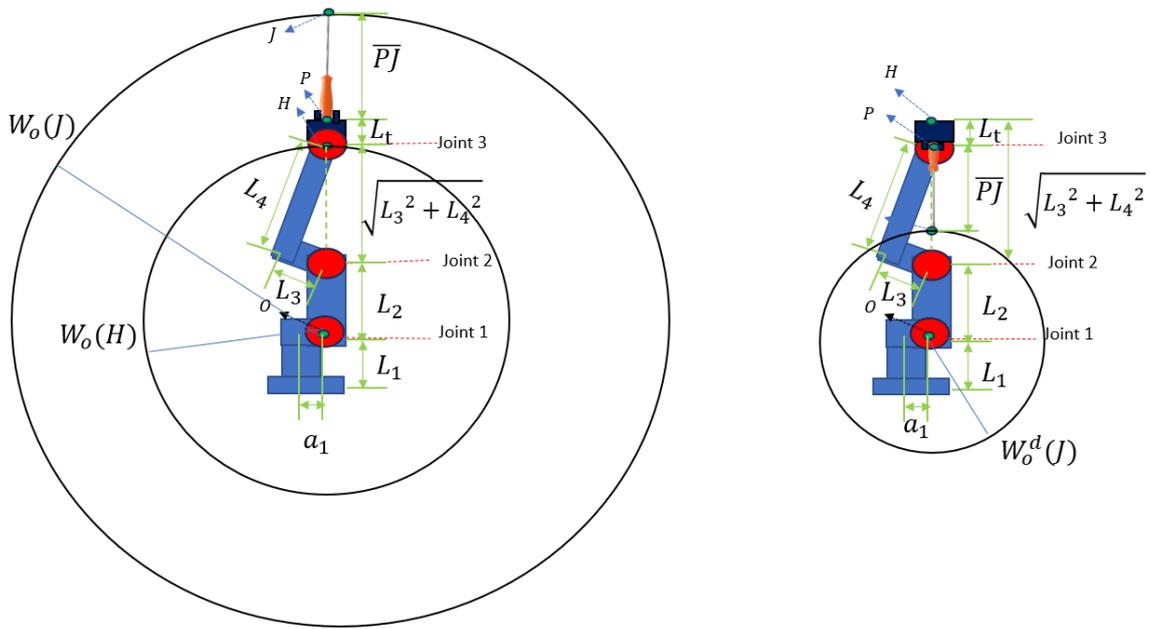

**Figure 13.** Workspace of an Ideal Elbow Manipulator.

It can be shown in the figure:

$$W_o^d(J) \subseteq W_o(H) \subseteq W_o(J). \tag{50}$$

For an ideal elbow manipulator, the radiuses of workspaces are expressed as below:

$$\text{Radius of } W_o(H) = L_2 + \sqrt{L_3^2 + L_4^2} \tag{51}$$

$$\text{Radius of } W_o(J) = L_2 + \sqrt{L_3^2 + L_4^2} + L_t + \overline{PJ} \tag{52}$$

$$\text{Radius of } W_o^d(J) = L_2 + \sqrt{L_3^2 + L_4^2} - L_t - \overline{PJ} \tag{53}$$

where $L_2$, $L_3$, and $L_4$ are links' length of the manipulator. $L_t$ is the length of the end-effector and $\overline{PJ}$ is the length of the object that mounts on the end-effector.

Because the camera and the tool must be able to rotate in all 3 DoFs when they are at the target position. So, the dexterous workspace is used as the working space of the visual system and the tool manipulation system. Dexterous workspace of the optical center is used for the visual system and Dexterous workspace of the far endpoint of the tool is used for the tool manipulator system.

$$\text{Radius of } W_o^d(Visual) = r_V = L_2 + \sqrt{L_3^2 + L_4^2} - L_t - \overline{PC}_{camera}$$
$$= 1.716 \ m \ \text{(From Specification Appendix Table A1)} \tag{54}$$

$$\text{Radius of } W_o^d(Tool) = r_T = L_2 + \sqrt{L_3^2 + L_4^2} - L_t - \overline{PJ}_{tool}$$
$$= 1.606 \ m \ \text{(From Specification Appendix Table A1)} \tag{55}$$

Figure 14 shows the workspace setup for the hybrid system. The dexterous space of robot arm with camera is a sphere with radius of $r_V$ and sphere center is $O_V$ and the dextrous space of the robot arm with tools is a sphere the radius is $r_T$ and center of this sphere is $O_T$. Two systems intersect with the ground with an angle $\theta_1$ and $\theta_2$.

$$\theta_1 = \arcsin\left(\frac{L_1}{r_V}\right) = 16.77° \tag{56}$$

$$\theta_2 = \arcsin\left(\frac{L_1}{r_T}\right) = 17.95° \tag{57}$$

To avoid interference of the visual and the tool manipulation systems (as the third geometric constraint of the operational space), the reachable workspaces (the maximum reach) of two systems should have no overlap. The reachable workspaces of two systems are calculated from Equation (51) is:

$$\text{Radius of } W_o(Visual) = r_{Vr} = L_2 + \sqrt{L_3^2 + L_4^2} + L_t + \overline{PJ}_{camera}$$
$$= 2.044 \ m \ \text{(From Specification Appendix Table A1)} \tag{58}$$

$$\text{Radius of } W_o(Tool) = r_{Tr} = L_2 + \sqrt{L_3^2 + L_4^2} + L_t + \overline{PJ}_{tool}$$
$$= 2.138 \ m \ \text{(From Specification Appendix Table A1)} \tag{59}$$

Let $L_{VT}$ is the distance between $O_V$ and $O_T$. The following relationship must be satisfied:

$$L_{VT} \geq r_{Vr} + r_{Tr} = 4.182 \ m \tag{60}$$

The visual system is able to detect the tool when it gets close to the target pose (Just above the bolt). Also reference points are put near the tool's target pose. Assume a marker is put at where the bolt is, and this marker represents as an approximated location of all interest points and reference points. Therefore, the marker should be within the dexterous space of the tool manipulation system. Set up a coordinate system with $\bar{O}_V$ (the projection of $O_V$ on the ground with $a_1$ offset) as the origin. The projection of $O_T$ on the ground with $a_1$ offset is $\bar{O}_T$. Assume the marker is placed on the line $\overline{\bar{O}_V \bar{O}_T}$. The distance m from marker M to $\bar{O}_V$ should be:

$$L_M \geq a_1 + L_{VT} - r_T \cos(\theta_2) \geq 2.829 \ m \tag{61}$$

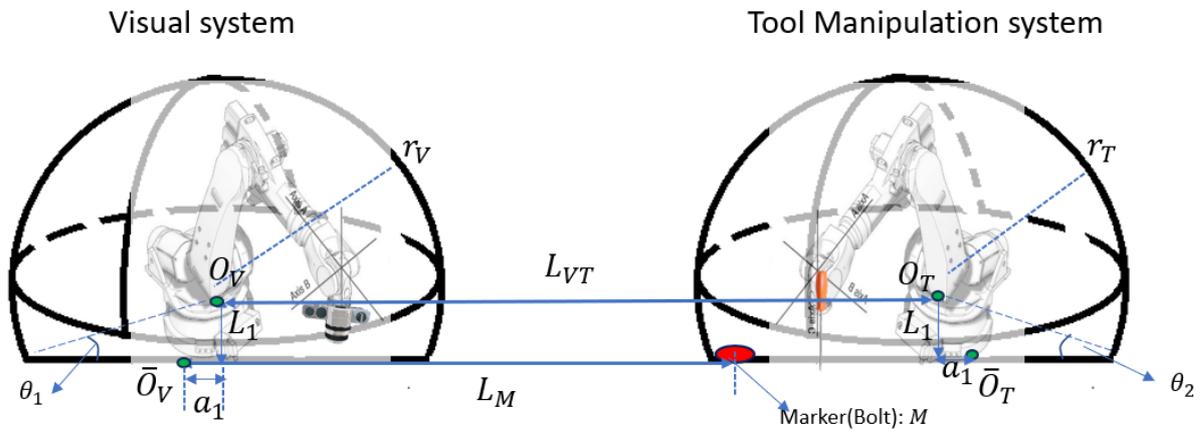

**Figure 14.** The workspace configuration of the multi-robotic system.

## 5.2. Detectable Space for the Stereo Camera

From Appendix Table A3, the angle of view in width and height for both lens in stereo camera is $\alpha = 86.05°$, and $\beta = 55.35°$. The detectable space for each lens can be modeled as the inner area of a cone with angle $\alpha$ and $\beta$. And the overlapped space is the detectable space for the stereo camera. The model is shown below (Figure 15). Point C is the center of the baseline whose length is denoted as $b$ or center of the camera system. The overlap area is also a cone with angles of vertex $\alpha$ and $\beta$ with the offset $d$ from baseline. Point Q is the vertex of the cone.

$$d = \frac{b/2}{\tan\left(\frac{\alpha}{2}\right)} = 0.064 m \tag{62}$$

There is an upper bound of the distance between the object to the camera center; if the object is too far away from the camera center, the dimensions of projected images are too small to be measured. Suppose to have a clear image of the fiducial markers (circle shapes), it is required that the diameter of the projected image should takes at least 5-pixel numbers in the image frame:

$$N_H \geq 5 \ \text{resolution} \tag{63}$$

where $N_H$ is the pixel number in the image frame.

Select the Zed camera's mode so that its resolution is 1920*1082 and the image sensor size is 5.23mm X 2.94mm. Then numbers resolution in unit length is:

$$n = 367 \ \text{resolution/mm} \tag{64}$$

Then the range of markers' diameter on the image frame is:

$$d_H = \frac{N_H}{n} \geq 0.0136 \ mm \tag{65}$$

Also assume in the inertial frame, the diameter of attached fiducial marks is:

$$D_H = 12 \ mm \tag{66}$$

Then from pinhole model of the camera, the range of distance between markers and camera center Z is:

$$Z = \frac{f_u D_H}{d_H} \leq 2.47 \ m = Z_{max} \tag{67}$$

where $Z_{max}$ is the maximum depth of camera to detect the markers. This parameter defines the height of the cone in Figure 15. The detectable space (abiding the second geometric constraint) forms an elliptic cone with different angles of vertex.

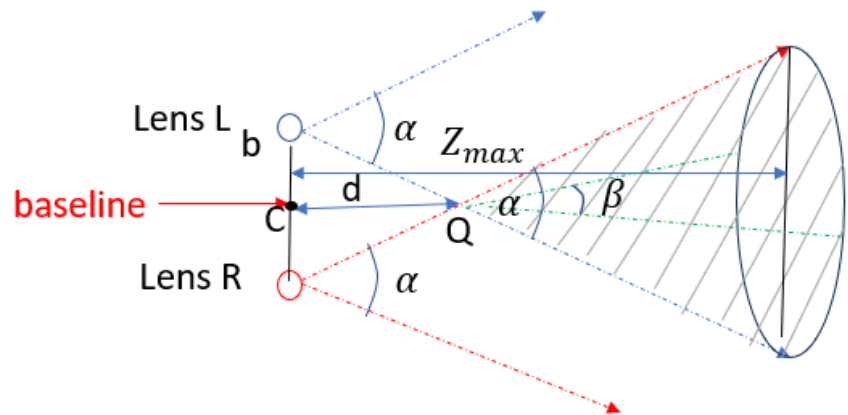

**Figure 15.** The Detectable Space of the Stereo Camera.

## 5.3. Camera Operational Space Development

In Figure 16, set the camera lens so that it is always parallel to the face of the marker and the baseline is parallel to $\bar{X}_V$ axis. In this case, lens is kept facing downward when detecting the marker on the ground. As discussed in the above section, the detectable space of the camera is modeled as a cone. This cone intersects with the horizontal plane and forms an oval. To have the camera detect the marker, the marker must be contained inside the oval.

An inertia coordinate system $\bar{O}_V \bar{X}_V \bar{Y}_V \bar{Z}_V$ is set up with its origin seated at the bottom center of the visual system. Make point $C$ as the location of the camera center and its coordinates in the inertial frame are $(X_C^V, Y_C^V, Z_C^V)$. $Q$ is the vertex of the detectable elliptic cone, and $Q'$ is the projection of $Q$ on the horizontal plane. $M$ is the marker.

The elliptic cone of the camera's detectable space always intersects with the horizontal plane. Consider an extreme case when point $C$ is located at the apex of the visual system workspace as shown in Figure 16. Then the $Z_C^V$ coordinate of $C$ at the inertial frame is:

$$Z_C^V\{max\} = L_1 + r_V = 2.211 \, m \leq Z_{max} \tag{68}$$

where $Z_{max}$ defined above is the maximum depth of camera to detect the markers. Therefore, the cone intersects with the horizontal plane when the camera center $C$ is placed at any location of the workspace's contour.

From expression of Cartesian coordinates in the inertia frame:

$$a_{Q'} = (Z_C^V - d)tan(\frac{\alpha}{2}), \ b_{Q'} = (Z_C^V - d)tan(\frac{\beta}{2}) \tag{69}$$

where $d$ is the same offset in Equation (62). $a_{Q'}$ and $b_{Q'}$ are major radius and minor radius of the oval. Also coordinates of points $Q'$ and $M$ are $(X_C^V, Y_C^V, 0)$ and $(L_M, 0, 0)$. $Q'$ is the center of the oval. To ensure the marker $M$ is inside the projected oval, it is required that:

$$\frac{(X_C^V - L_M)^2}{a_{Q'}^2} + \frac{(Y_C^V)^2}{b_{Q'}^2} \leq 1 \tag{70}$$

Plug Equation (69) into Equation (70):

$$\frac{(X_C^V - L_M)^2}{tan(\frac{\alpha}{2})^2} + \frac{(Y_C^V)^2}{tan(\frac{\beta}{2})^2} \leq (Z_C^V - d)^2 \tag{71}$$

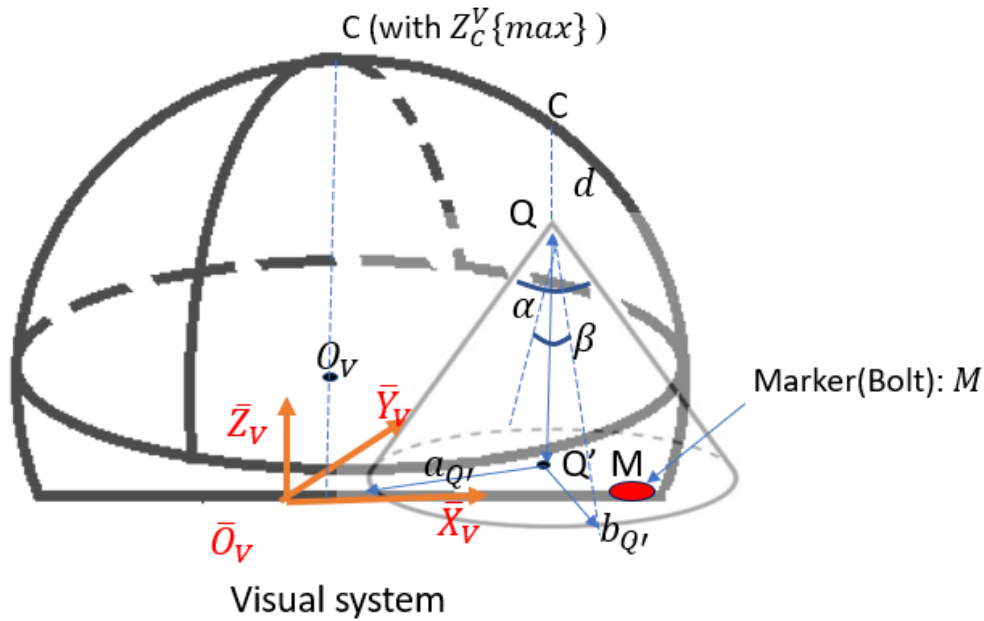

**Figure 16.** The Detectable Elliptic Cone in the Inertial Visual System Frame.

Inequality (71) is exactly a mathematic expression of all points $C(X_C^V, Y_C^V, Z_C^V)$ within a cone whose opening is in the positive Z direction with its vertex at $(L_M, 0, d)$ and the opening parameters as $tan(\frac{\alpha}{2})$=0.934 and $tan(\frac{\beta}{2})$=0.524. Therefore, this inequality defines the second geometric constraint of the camera center. It forms an elliptical cone with its vertex located in Figure 17 at point E, which is the offset point of the marker $M$ from the ground. As $C(X_C^V, Y_C^V, Z_C^V)$ should also be within the sphere of the workspace (abiding the first geometric constraint), the camera operational space is presented as the overlap between the cone and the sphere (the shaded area in Figure 17).

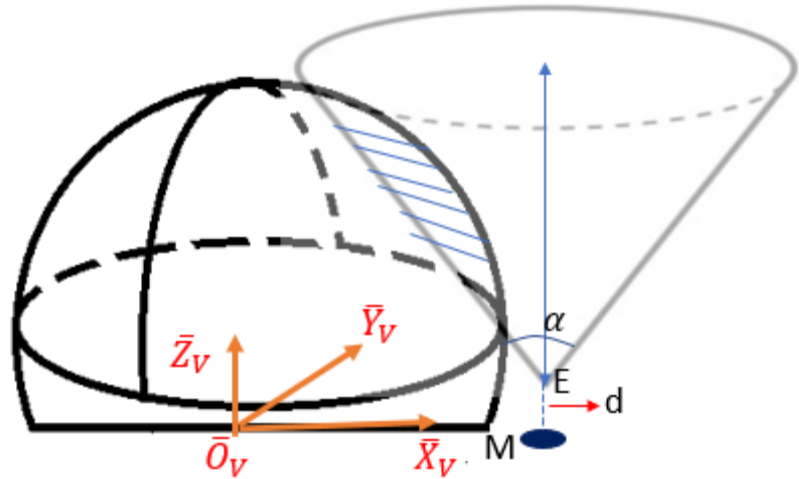

**Figure 17.** Illustration of Camera Operational Space (the shaded area).

From Figure 17, the marker should not be too far from the visual system, otherwise there is no overlap region between the cone and the sphere. The largest value of $L_M$, which is the distance from the marker to the coordinate center, occurs when the cone is tangent to the sphere as shown in Figure 18. To find the limit of $L_M$, draw a horizontal line across point $O_V$ so that it intersects the surface of the cone at point A and the center line of the cone at point B as shown in Figure 18. Let $\overline{O_V A}$ and $\overline{AB}$ are line segments' length.

Therefore, from the geometry, the upper limit of $L_M$ is:

$$L_M < \overline{O_V A} + \overline{AB} + a_1 = \frac{r_V}{\cos\left(\frac{\alpha}{2}\right)} + (L_1 - d)\tan\left(\frac{\alpha}{2}\right) + a_1 \tag{72}$$

Combining with the lower bound of $L_M$ in Equation (61), the marker M should be placed in the system so that:

$$2.829\ m \leq L_{VT} - r_T\cos(\theta_2) \leq L_M < \frac{r_V}{\cos\left(\frac{\alpha}{2}\right)} + (L_1 - d)\tan\left(\frac{\alpha}{2}\right) + a_1 = 2.925\ m \tag{73}$$

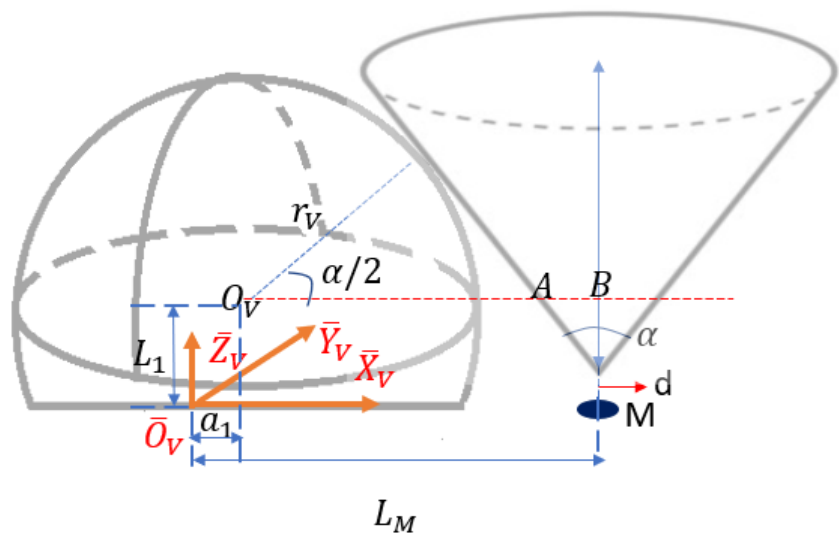

**Figure 18.** An Extreme Case of Marker Position.

### 5.4. Mathematic Expression for Node Coordinates within the Camera Operational Space

From Equation (73), the closer the marker's distance $L_M$ to its lower bound, the larger the operational space is. In general, larger operational space is preferred, because the camera has a larger space for searching the optimal location. Set $L_M$ equals to its lower bound:

$$L_M = 2.83m \tag{74}$$

Get the cross-section area at $\bar{X}_V - \bar{Z}_V$ plane (Figure 19).

The function of semi-circle and the line EFG is:

$$(\bar{Z}_V - L_1)^2 + (\bar{X}_V - a_1)^2 = r_V^2 \tag{75}$$

$$\bar{X}_V = -tan(\frac{\alpha}{2})(\bar{Z}_V - d) + L_M \tag{76}$$

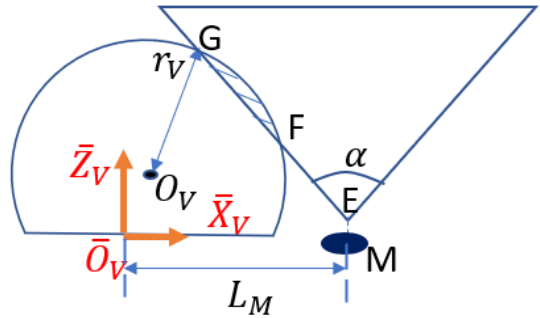

**Figure 19.** The Camera Operational Space Cross-section at $\bar{X}_V - \bar{Z}_V$ plane.

Combine Equations (75) and (76) and solve:

$$(1 + tan(\frac{\alpha}{2})^2)\bar{Z}_V^2 - 2(tan(\frac{\alpha}{2})^2 d + tan(\frac{\alpha}{2})(L_M - a_1) + L_1)\bar{Z}_V + tan(\frac{\alpha}{2})^2 d^2 + 2tan(\frac{\alpha}{2})(L_M - a_1)d + (L_M - a_1)^2 - r_V^2 + L_1^2 = 0 \tag{77}$$

Solve Equation (77) by plugging numbers:

$$1.872\,\bar{Z}_V^2 - 6.061\bar{Z}_V + 4.670 = 0 \tag{78}$$

Solve: $\bar{Z}_{V_{max}}, \bar{Z}_{V_{min}} = 1.972\,m,\ 1.265\,m \tag{79}$

$\bar{Z}_{V_{max}}$ and $\bar{Z}_{V_{min}}$ are the largest and smallest $\bar{Z}_V$ coordinates of the camera center $C$ that is within the camera operation space.

Take a value of $Z_C^V$ ($\bar{Z}_V$ coordinates of the camera center $C$) in the range $(\bar{Z}_{V_{min}}, \bar{Z}_{V_{max}}) = (1.265\,m, 1.972\,m)$. Set a plane $\bar{Z}_V = Z_C^V$, and that plane intersects the camera operational space and forms a shade area, which is the overlap between the circle and the oval, as shown in Figure 14. The inequality governing points $C$ within sphere at a specific $Z_C^V$ is:

$$(\bar{X}_V - a_1)^2 + \bar{Y}_V^2 \leq r_V^2 - (Z_C^V - L_1)^2 \tag{80}$$

The inequality governing points $C$ within cone at $Z_C^V$ is:

$$\frac{(\bar{X}_V - L_M)^2}{tan(\frac{\alpha}{2})^2} + \frac{(\bar{Y}_V)^2}{tan(\frac{\beta}{2})^2} \leq (Z_C^V - d)^2 \tag{81}$$

By plotting those inequalities in Figure 20, the shaded area is where $X_C^V$ and $Y_C^V$ ($\bar{X}_V$ and $\bar{Y}_V$ coordinates of the camera center $C$) should be located within. The intersection of the two curves occurs when equality holds for (80) and (81), that is:

$$(\bar{X}_V - a_1)^2 + \bar{Y}_V^2 = r_V^2 - (Z_C^V - L_1)^2 \tag{82}$$

$$\frac{(\bar{X}_V - L_M)^2}{\tan(\frac{\alpha}{2})^2} + \frac{(\bar{Y}_V)^2}{\tan(\frac{\beta}{2})^2} = (Z_C^V - d)^2 \tag{83}$$

Solve by plugging numbers:

$$(\bar{Y}_{V_{max}}, \ \bar{Y}_{V_{min}}) = \pm 0.5 *$$
$$\sqrt{3.44 * Z_C^{V2} - 2.0241 * Z_C^V - 29.802 + 2.451 * \sqrt{-29.475 * Z_C^{V2} + 23.707 * Z_C^V + 137.28}} \tag{84}$$

$\bar{Y}_{V_{max}}$ and $\bar{Y}_{V_{min}}$ are the largest and the smallest $\bar{Y}_V$ coordinates of the camera center $C$ in the camera operation space with specific $Z_C^V$.

Take $Y_C^V$ in the range $(\bar{Y}_{V_{min}}, \ \bar{Y}_{V_{max}})$, then for specific $X_C^V$:

$$\bar{X}_{V_{max}} = \sqrt{r_V^2 - (Z_C^V - L_1)^2 - Y_C^{V2}} + a_1 \tag{85}$$

$$\bar{X}_{V_{min}} = L_M - \sqrt{(Z_C^V - d)^2 \tan(\frac{\alpha}{2})^2 - Y_C^{V2} \frac{\tan(\frac{\alpha}{2})^2}{\tan(\frac{\beta}{2})^2}} \tag{86}$$

$\bar{X}_{V_{max}}$ and $\bar{X}_{V_{min}}$ are the largest and the smallest $\bar{X}_V$ coordinates of the camera center $C$ in the camera operation space with specific $Z_C^V$ and $Y_C^V$.

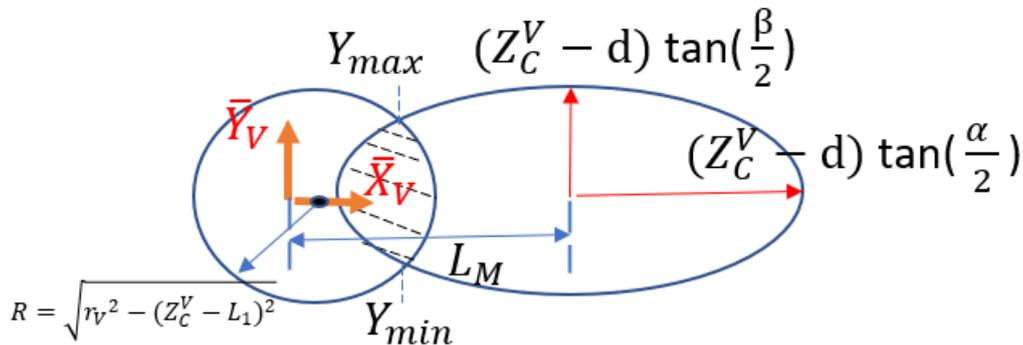

**Figure 20.** The Camera Operational Space Cross-section at $\bar{X}_V - \bar{Y}_V$ plane.

## 5.5. Numerical Solution of the Ideal Camera Operation Space

With known parameters from the camera and robot arm specifications, all possible location of the camera center $C(X_C^V, Y_C^V, Z_C^V)$ inside the camera operational space can be derived from the following steps:

1. Calculate $\bar{Z}_{V_{max}}$ and $\bar{Z}_{V_{min}}$ by Equation (77), and mesh grid $Z_C^V$ in $(\bar{Z}_{V_{min}}, \bar{Z}_{V_{max}})$.

2. Take a mesh value $Z_C^V$ and calculate $\bar{Y}_{V_{max}}$ and $\bar{Y}_{V_{min}}$ by Equations (82) and (83), and mesh grid $Y_C^V$ in $(\bar{Y}_{V_{min}}, \bar{Y}_{V_{max}})$.

3. Take a mesh $Y_C^V$ and calculate $\bar{X}_{V_{max}}$ and $\bar{X}_{V_{min}}$ by Equations (85) and (86), and mesh grid $X_C^V$ in $(\bar{X}_{V_{min}}, \bar{X}_{V_{max}})$.

Each above computed $C(X_C^V, Y_C^V, Z_C^V)$ is inside the camera operation space. And those three steps completely account for all geometric constraints defined above. The number of mesh points (nodes available for searching) depends on the mesh grid size.

## 5.6. The Camera Operational Space from the Realistic Robot Manipulator

The above procedures of finding the camera operational space assumes the use of ideal robotic manipulators, whose joints are freely to move from $[0, 2\pi]$. However, for realistic robotic manipulators, their revolutionary movement is limited within certain ranges of angles, and the ranges of motion in every joint of the specific robot manipulator: ABB IRB 4600 are listed in Appendix Table A2.

The geometric shape of the operational space for a realistic robotic manipulator is usually irregular (unlike an ideal manipulator whose operational space is the combination of a sphere and a cone) and the mathematical methods of directly finding the nodes within the space is computationally expensive. A faster way of generating realistic camera operational space is to decrease the ideal operational space by taking out nodes that are out of the operational ranges. Those outlier nodes can be found from the inverse kinematic model of the robotic manipulator. Figure 21 shows the decreased nodes from the ideal camera operational space to the realistic space.

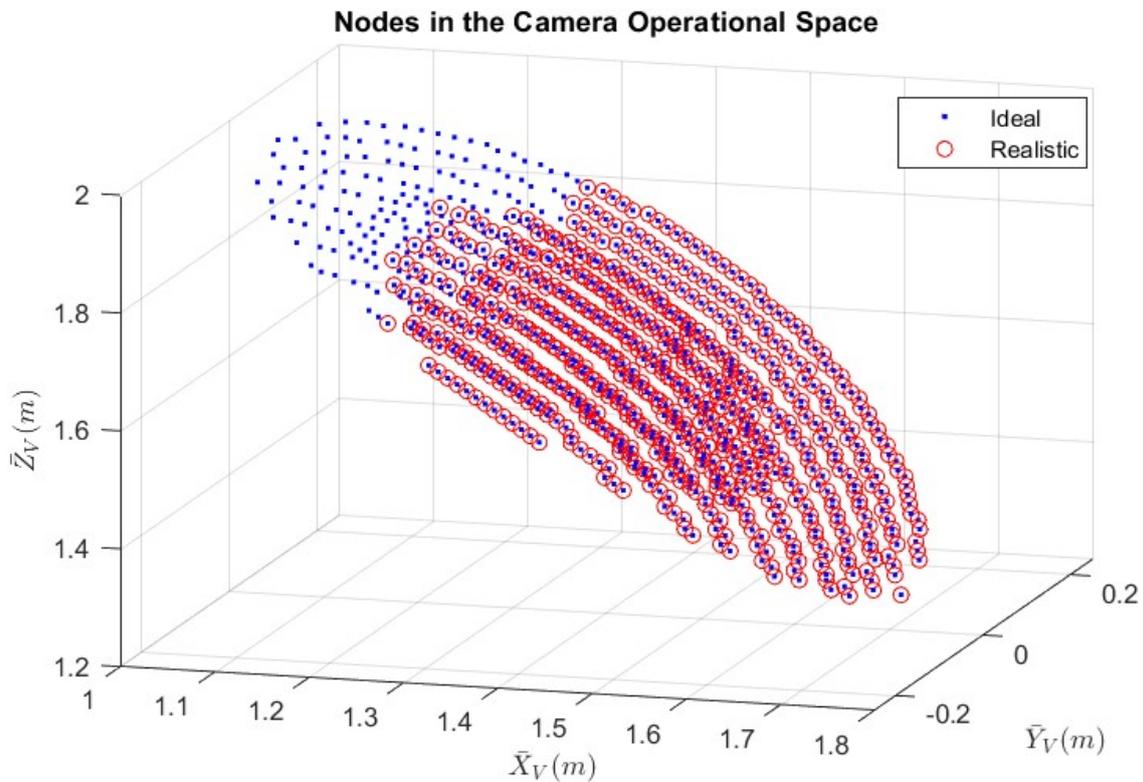

**Figure 21.** Reduction of Nodes from the ideal to the realistic camera operational space.

## 6. Simulation Results

A simulation in MATLAB is used to validate the proposed algorithm. In the simulation, the environment is set up with two minimums in the space (only one is the global minimum). $K_{est}$ and $K_{sd}$ are tunable parameters and set as $K_{est} = 5, and\ K_{sd} = 50\ (m^{-1})$. The initial available energy bound is set to be $E_{bound} = 12$ (Watt Second (ws) or Joule). Figure 22 shows the sequences of nodes being explored in iteration steps by setting different energy threshold $E_T$.

By trials of different simulation runs, it can be concluded that:

- When $E_T$ is large (in this scenario, $3ws < E_T \leq 12ws$), the algorithm is conservative, so that it has only searched a small area that excludes the global minimum.
- When $E_T$ is in mid-range (in this scenario, $0.5ws \leq E_T \leq 3ws$), the algorithm drives the camera to the global minimum.

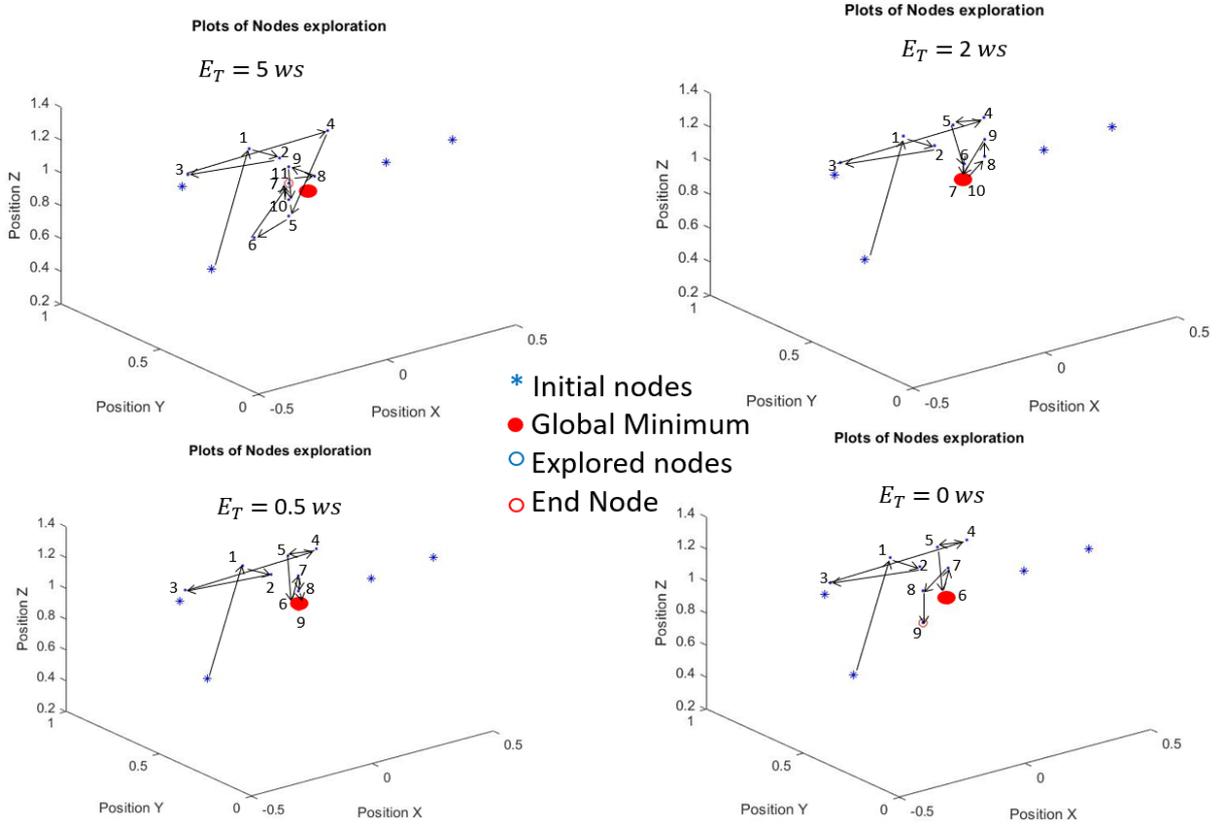

**Figure 22.** Plots of sequential exploration with different $E_T$.

When $E_T$ is small (in this scenario, $0ws \leq E_T < 0.5ws$), the algorithm is aggressive. Although it has searched a large amount of areas that includes the global minimum, it doesn't have enough energy to move the camera back to the best location explored in the previous stages

The two tunable parameters $K_{est}$ and $K_{sd}$ are selected randomly in the simulation. Those parameters are used in the models to estimate the noise spatial distribution in the environment. As stated in section 4, the positive parameter $K_{est}$ indicates that the weight of an explored node in the estimation function is based on its Euclidean distance from the node to be estimated. Larger $K_{est}$ in Equation (4) implies that Euclidean distance has smaller influence on the weight. In other words, with increasing $K_{est}$, the estimation result of an unexplored node is more different than that of an explored node nearby. As a result, the algorithm is more likely to search nodes around the local minimum. Similarly, $K_{sd}$ is proportional to the magnitude of the standard deviation in the stochastic modified process. Smaller $K_{sd}$ causes smaller standard deviation and thus more deterministic evaluation of an unexplored node, which discourages the algorithm to explore areas that are away from the local minimum.

Therefore, by increasing $K_{est}$ or decreasing $K_{sd}$, we should expect that the algorithm has higher tendency to exploit over explore. In this paper, we have developed a parameter to present the degree of exploration in a simulation by measuring the average distance between the new explored node to the nearest previously explored node.

$$Avg\_Dis_{new} = \frac{\sum_{i=1}^{M} \min_{j \in S_{index-exp}} (\overline{ED}(P_j^{exp}, P_i^{new}))}{M} \tag{87}$$

Where $M$ is the total number of iterations in a simulation, $P_i^{new}$ is the position of the new explored node at $i^{th}$ iteration.

$$S_{index-exp} = \{index | Node_{index} \in S_{explored}\}, \tag{88}$$

$$\overline{ED}(P_a, P_b) = \sqrt{(P_{ax} - P_{bx})^2 + (P_{ay} - P_{by})^2 + (P_{az} - P_{bz})^2},$$

with $P_a, P_b$ are 3D locations of $Node_a$ and $Node_b$:  (89)

$$P_a = (P_{ax}, P_{ay}, P_{az}) \ and \ P_b = (P_{bx}, P_{by}, P_{bz})$$

In a simulation, a large value of the parameter $Avg\_Dis_{new}$ means new explored nodes in each search iteration are generally far away from previously explored nodes, and the algorithm succeeds in exploring a large area. A sensitivity test is given by varying $K_{est}$ and $K_{sd}$ to show how these two parameters affect the searching process in the same simulation environment. Table 2 presents the sensitivity analysis of $K_{est}$ by keeping a constant $K_{sd} = 50 \ (m^{-1})$ while Table 3 presents the sensitivity analysis of $K_{sd}$ by keeping a constant $K_{est} = 50$ (No unit). All analysis tests are based on the same scenario in figure 22 with $E_{bound} = 20$ (ws) and $E_T = 2$ (ws).

**Table 2.** Sensitivity Analysis with Varying $K_{est}$. ($K_{sd} = 50 \ (m^{-1})$)

| $K_{est}$ | $Iteration \ Numbers$ | $Avg\_Dis_{new} \ (m)$ | Stop at global Minimum |
|---|---|---|---|
| 90 | 93 | 0.059 | NO |
| 70 | 83 | 0.066 | NO |
| 50 | 67 | 0.073 | YES |
| 30 | 46 | 0.085 | YES |
| 10 | 38 | 0.097 | YES |
| 5 | 47 | 0.102 | YES |
| 1 | 20 | 0.131 | NO |

**Table 3.** Sensitivity Analysis with Varying $K_{sd}$. ($K_{est} = 50$)

| $K_{sd}(m^{-1})$ | $Iteration \ Numbers$ | $Avg\_Dis_{new} \ (m)$ | Stop at global Minimum |
|---|---|---|---|
| 90 | 36 | 0.086 | NO |
| 70 | 62 | 0.079 | YES |
| 50 | 67 | 0.073 | YES |
| 30 | 84 | 0.066 | YES |
| 10 | 127 | 0.060 | YES |
| 5 | 176 | 0.046 | YES |
| 1 | 264 | 0.037 | NO |

Table 2 and Table 3 show that by decreasing $K_{est}$ or increasing $K_{sd}$, the algorithm explores more areas as indicated by the parameter $Avg\_Dis_{new}$. The iteration numbers reduce as more areas have been explored because fewer nodes can be reached when the total energy for moving is bounded. Those conclusions can be derived from the analysis. On one hand, the algorithm is too conservative and searches few areas when $K_{est}$ is too large or $K_{sd}$ is too small. On the other hand, the algorithm is too aggressive and skips exploiting when $K_{est}$ is too small or $K_{sd}$ is too large. Both cases will make the node not end up with the global minimum. Therefore, a moderate combination of $K_{est}$ and $K_{sd}$ is preferred. Future research can focus on development of adaptive algorithms that tune $K_{est}$ and $K_{sd}$ over iterations based on errors between real measurements and estimations.

In addition, the numerical values of $E_T$ are also picked randomly for testing. In real applications, a fixed $E_T$ should be selected before the operation of this algorithm. The selection of $E_T$ is related to the specific application scenario, size of camera's operational space, and total energy available $E_{bound}$ at beginning. However, the general thumb of the rule is the following: choose small values of $E_T$ when both $E_{bound}$ and operational space is large to encourage exploring over exploiting; otherwise, choose large $E_T$.

## 7. Conclusions

In this article, we have developed an algorithm that a camera can explore the workspace in a manufacturing environment and search for a location so that images' noise is minimized within the space that it can reach. This article also provides detailed development of the camera's operational space for a specific application. Results in a virtual environment have shown the algorithm succeed in bringing the camera to the optimal (or suboptimal if the optimal one is unreachable) position in this specific scenario. By reducing processing time of images, this algorithm can be used for various visual servoing applications in high-speed manufacturing.

However, challenges arise in some possible scenarios of real applications. For instance, if environmental factors vary too rapidly across space, a fine gridding of the space is required to manifest those variations in the estimation function, which increases computational complexity, resulting in poor real-time performances. The same issues exist in cases where the operational space of the camera is too large.

Other limitations include that this algorithm is incapable of differentiating contrast in images at different locations. Low contrast other than image noises occurs as another issue in visual servoing environment; for instance, low contrast in soft tissues has been shown negative effects on the performance of visual servoing controller in medical applications [38]. In addition, this algorithm does not account for cases when object is partially obstructed, which is very common in manufacturing environment. Limitations in the camera hardware that cannot account for additive noise are also not addressed in this paper. Those effects cannot be eliminated by any image denoising processes. For example, sensors with lower dynamic range may produce images where details are lost in extreme lighting conditions (bright sunlight or deep shadows). Another example is that some lenses may cause vignetting (darkening around the edges) or distortion, which can degrade the perceived image quality.

Simulation results show how energy threshold $E_T$ affects the behavior of the algorithm. A moderate threshold is suggested so that the algorithm can search for enough amount of area but enables to drive the camera back to the minimum with limited energy available. The analysis of $E_T$'s value in different applications is a focus of the future work. Future work may also include adaptive algorithms that update $K_{est}$ and $K_{sd}$ online and an adaptive function for energy estimation. Also, orientations of the camera are fixed in this paper. In the future, we would like to develop advanced algorithms that provide not only the optimal of positions but also the orientations inside the space.

In summary, this paper initially explores an adaptive algorithm of searching the optimal location in space with respective to image noises for observation in ETH or ETH/EIH cooperative configurations. Although several improvements of this algorithm can be addressed in the future, it has already shown a great potential to be applied as an upgraded feature in some of the real manufacturing tasks.

## Appendix A

In this section, we will show the geometric model of a specific robot manipulator ABB IRB 4600 45/2.05[23] and a figure of a camera model: Zed 2 with dimensions [24]. This section also contains specification tables of robots' dimensions, camera, and motor installed inside the joints of manipulators.

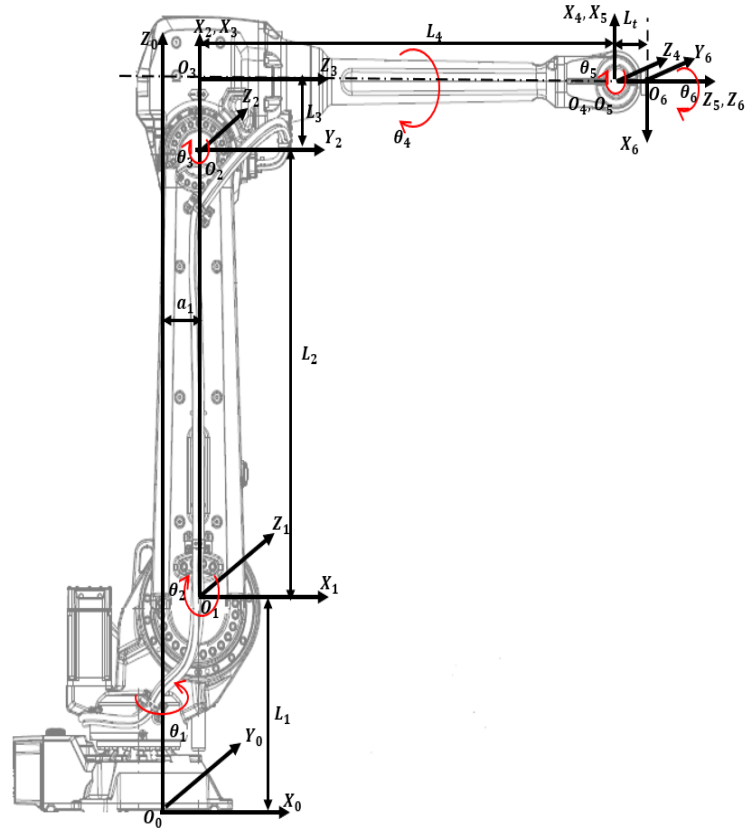

**Figure A1.** IRB ABB 4600 Model with attached frames.

*Dimensions are in mm*

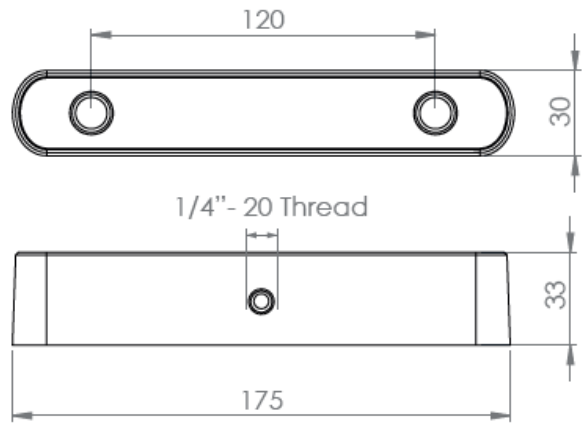

**Figure A2.** Zed 2 stereo camera model with dimensions.

**Table A1.** Specification Table of ABB IRB 4600 45/2.05 Model (Dimensions).

| Parameters | Values |
|---|---|
| Length of Link 1: $L_1$ | 495 mm |
| Length of Link 2: $L_2$ | 900 mm |
| Length of Link 3: $L_3$ | 175 mm |
| Length of Link 3: $L_4$ | 960 mm |
| Length of Link 1 offset: $a_1$ | 175 mm |
| Length of Spherical wrist: $L_t$ | 135 mm |
| Tool length (screwdriver): $\overline{PJ}_{tool}$ | 127 mm |

**Table A2.** Specification Table of ABB IRB 4600 45/2.05 Model (Axis Working range).

| Axis Movement | Working range |
|---|---|
| Axis 1 rotation | +180° to -180° |
| Axis 2 arm | +150° to -90° |
| Axis 3 arm | +75° to -180° |
| Axis 4 wrist | +400° to -400° |
| Axis 5 bend | +120° to -125° |
| Axis 6 turn | +400° to -400° |

**Table A3.** Specification Table of Stereo Camera Zed 2.

| Parameters | Values |
|---|---|
| Focus length: f | 2.8 mm |
| Baseline: B | 120 mm |
| Weight: W | 170g |
| Depth range: | 0.5m-25m |
| Diagonal Sensor Size: | 6mm |
| Sensor Format: | 16:9 |
| Sensor Size: W X H | 5.23mm X 2.94mm |
| Resolution | 1920 pixels X 1080 pixels |
| Angle of view in width: $\alpha$ | 86.09° |
| Angle of view in height: $\beta$ | 55.35° |

**Table A4.** Specification Table of Motors and gears.

| Parameters | Values |
|---|---|
| **DC Motor** | |
| Armature Resistance: $R$ | 0.03 Ω |
| Armature Inductance: $L$ | 0.1 mH |
| Back emf Constant: $K_b$ | 7 mv/rpm |
| Torque Constant: $K_m$ | 0.0674 N/A |
| Armature Moment of Inertia: $J_a$ | 0.09847 kg$m^2$ |
| **Gear** | |
| Gear ratio: $r$ | 200:1 |
| Moment of Inertia: $J_g$ | 0.05 kg$m^2$ |
| Damping ratio: $B_m$ | 0.06 |

## Appendix B

In this section, we will show forward kinematics and inverse kinematics of the 6 DoFs revolute robot manipulators. The results are consistent with the model ABB IRB 4600.

Forward kinematics refers to the use of kinematic equations of a robot to compute the position of the end-effector from specified values for the joint angles and parameters. The Equations (B1)-(B5) are summarized in the below:

$$n_x = c_1 s_{23}(s_4 s_6 - c_4 c_5 c_6) - s_1(s_4 c_5 c_6 + c_4 s_6) - c_1 c_{23} s_5 c_6$$
$$n_y = s_1 s_{23}(s_4 s_6 - c_4 c_5 c_6) + c_1(s_4 c_5 c_6 + c_4 s_6) - s_1 c_{23} s_5 c_6$$
$$n_z = c_{23}(s_4 s_6 - c_4 c_5 c_6) + s_{23} s_5 c_6 \tag{B1}$$

$$s_x = c_1 s_{23}(s_4 c_6 + c_4 c_5 c_6) + s_1(s_4 c_5 s_6 - c_4 c_6) + c_1 c_{23} s_5 s_6$$
$$s_y = s_1 s_{23}(s_4 c_6 + c_4 c_5 c_6) - c_1(s_4 c_5 c_6 - c_4 c_6) + s_1 c_{23} s_5 s_6$$
$$s_z = c_{23}(s_4 c_6 + c_4 c_5 c_6) - s_{23} s_5 s_6 \tag{B2}$$

$$a_x = -c_1 s_{23} c_4 s_5 - s_1 s_4 s_5 + c_1 c_{23} c_5$$
$$a_y = -s_1 s_{23} c_4 s_5 + c_1 s_4 s_5 + s_1 c_{23} c_5$$
$$a_z = c_{23} c_4 s_5 - s_{23} c_5 \tag{B3}$$

$$d_x = L_t(-c_1 s_{23} c_4 s_5 - s_1 s_4 s_5 + c_1 c_{23} c_5) + c_1(L_3 s_{23} + L_2 s_2 + a_1)$$
$$d_y = L_t(-s_1 s_{23} c_4 s_5 + c_1 s_4 s_5 + s_1 c_{23} c_5) + s_1(L_3 s_{23} + L_2 s_2 + a_1)$$
$$d_z = L_t(c_{23} c_4 s_5 - s_{23} c_5) + L_3 c_{23} + L_2 c_2 + L_1 \tag{B4}$$

$$\text{Note: } c_i \equiv \cos(q_i), \; s_i \equiv \sin(q_i)$$
$$c_{i,j} \equiv \cos(q_i + q_j), \; s_{i,j} \equiv \sin(q_i + q_j)$$
$$i, j \in \{1,2,3,4,5,6\} \tag{B5}$$

where $[n_x, n_y, n_z]^T$, $[s_x, s_y, s_z]^T$ and $[a_x, a_y, a_z]^T$ are the end-effector's directional vector of Yaw, Pitch and Roll in base frame $O_0 X_0 Y_0 Z_0$ (Figure A1). And $[d_x, d_y, d_z]^T$ are the vector of absolute position of the center of the end-effector in base frame $O_0 X_0 Y_0 Z_0$. For a specific model ABB IRB 4600-45/2.05 (Handling capacity: 45 kg/ Reach 2.05m) the dimensions and mass are summarized in Table A1.

Inverse kinematics refers to the mathematical process of calculating the variable joint angles needed to place the end-effector in a given position and orientation relative to the inertial base frame. The Equations (B6) – (B13) are summarized in the below:

$$p_x = d_x - L_t a_x$$
$$p_y = d_y - L_t a_y$$
$$p_z = d_z - L_t a_z \tag{B6}$$

$$q_1 = arctan(\frac{p_y}{p_x}) \tag{B7}$$

$$q_2 = \frac{pi}{2} - \arccos\left(\frac{L_2{}^2 + \left(\sqrt{p_x{}^2 + p_y{}^2} - a_1\right)^2 + (p_z - L_1)^2 - L_3{}^2 - L_4{}^2}{2L_2\sqrt{L_3{}^2 + L_4{}^2}}\right)$$
$$- arctan\left(\frac{p_z - L_1}{\sqrt{p_x{}^2 + p_y{}^2} - a_1}\right) \tag{B8}$$

$$q_3 = \pi - \arccos\left(\frac{L_2{}^2 + L_3{}^2 + L_4{}^2 - \left(\sqrt{p_x{}^2 + p_y{}^2} - a_1\right)^2 - (p_z - L_1)^2}{2L_2\sqrt{L_3{}^2 + L_4{}^2}}\right) - \arctan\left(\frac{L_4}{L_3}\right) \tag{B9}$$

$$q_5 = \arccos\left(c_1 c_{23} a_x + s_1 c_{23} a_y - s_{23} a_z\right) \tag{B10}$$

$$q_4 = \arctan\left(\frac{s_1 a_x - c_1 a_y}{c_1 s_{23} a_x + s_1 s_{23} a_y + c_{23} a_z}\right) \tag{B11}$$

$$q_6 = -\arctan\left(\frac{c_1 c_{23} s_x + s_1 c_{23} s_y - s_{23} s_z}{c_1 c_{23} n_x + s_1 c_{23} n_y - s_{23} n_z}\right) \tag{B12}$$

$$\text{Note: } c_i \equiv \cos(q_i), \; s_i \equiv \sin(q_i)$$
$$c_{i,j} \equiv \cos(q_i + q_j), \; s_{i,j} \equiv \sin(q_i + q_j) \tag{B13}$$

$$i, j \in \{1,2,3,4,5,6\}$$

where $[n_x, n_y, n_z]^T$, $[s_x, s_y, s_z]^T$, $[a_x, a_y, a_z]^T$ and $[d_x, d_y, d_z]^T$ have been defined above in the forward kinematic discussion.